\documentclass[letterpaper,11pt]{article}

\usepackage{jheppub} 
\usepackage{slashed}
\usepackage{graphicx}
\usepackage{subfig}
\usepackage{textcomp}
\usepackage{color}
\usepackage{wrapfig}
\usepackage{amsmath,amssymb}
\usepackage{feynmf}
\usepackage{footnote}
\usepackage{multirow}

\newcommand{\beq}{\begin{equation}}
\newcommand{\eeq}{\end{equation}}
\newcommand{\bea}{\begin{eqnarray}}
\newcommand{\eea}{\end{eqnarray}}

\graphicspath{{Figures/}}

\title{Probing top-antitop resonances with $t\bar{t}$ scattering 
at LHC14}

\author[a,b]{Da Liu,}
\author[b]{Rakhi Mahbubani}

\affiliation[a]{State Key Laboratory of Theoretical Physics, Institute of
Theoretical Physics, Chinese Academy of Sciences, Beijing, People's Republic of China.}
\affiliation[b]{Institut de Th\'eorie des Ph\'enom\`enes Physiques, EPFL, Lausanne, Switzerland.}
\abstract{
We explore the sensitivity of the LHC at 14 TeV centre-of-mass energy
(LHC14) to the single production and decay of top-antitop resonances
in the four-top final state.  We focus on the same-sign dilepton
channel, and work within a simplified model with a vector
boson coupling to the Standard Model only via its interactions with right-handed
top quarks.  We find it is possible to discover (exclude) such a vector boson 
with 300 fb$^{-1}$ of integrated luminosity up to a mass of 1.2 (1.6)
TeV for a modest coupling to tops of $g_\rho$=2.  We present our results as
an exclusion limit on the cross-section$\times$branching ratio for
ease of recasting, and
interpret them in the context of the gauge-singlet vector boson
$\rho_X$ present in many simple Composite Higgs theories.}

\begin{document}

\maketitle

\section{Introduction}

There are many compelling reasons to search for new physics coupling
to top quarks. By virtue of a large yukawa coupling, which is
  responsible for its 
electroweak-scale mass, the top quark contributes the largest quadratically divergent
contribution to the Higgs mass.  This intimate association with the
electroweak symmetry-breaking scale makes it plausible that the top is
also closely linked to whatever new physics makes the electroweak
scale natural.  Moreover, its relatively recent discovery means that its nature and
properties have not yet been explored in great detail. This is
particularly true of the right-handed (RH) top quark.  

One common feature of many potential solutions to the
  electroweak hierarchy problem is the
presence of new coloured partners for the top quark, that cancel its problematic
contribution to the higgs mass.  The large production cross sections of these
top partners, and their coloured relations, at the LHC, result in
uncomfortably strong constraints on their masses from recent null
searches. Limits on these top partners were around 700 GeV at the end
of Run 1 and are expected to fast approach the 1
TeV mark with Run 2 data.  Aside from naturalness, however, there seems
little reason to believe these coloured states to be the lighter than
any others in the particle spectrum of many leading natural UV
completions to the Standard Model (SM).  On the contrary, one might expect
their Renormalization Group running due to color to push them to
larger masses than uncoloured states.

This raises the question of whether current search strategies cast a
sufficiently wide net over this uncoloured theory space, or if there
are some interesting regions that might be overlooked.  One interesting example is a gauge-singlet vector that couples
dominantly to top quarks.  Such a particle is a robust feature
of any (economical)
Composite Higgs models where the right-handed top quark is
fully-composite singlet of the unbroken global symmetry of the strong
sector\footnote{In this limit, the explicit breaking of the symmetry
  protecting the mass of the Higgs will originate from a linear mixing
  of the third family doublet $(t_L,b_L)$ with the strong sector. As a result obtaining a light Higgs will be easier. }.  As a gauge singlet, this vector boson is constrained neither by precision
electroweak measurements, nor flavour physics,\footnote{Provided one
  implements a flavour story that forbids its couplings to light
  up-type quarks.}, and hence could be lighter than
all other composite states in the theory.  Its coupling to all other
SM particles, however, are suppressed by small mixings.

The leading tree-level production for this vector singlet is resonant
production in association with a top-antitop pair.  Current
resonance searches in the four-top final state, however, are tailored
to pair-produced resonances, the kinematics of which differs
significantly from our scenario.  A dedicated search may be necessary,
in order to improve the sensitivity for singly-produced resonances,
especially in the low mass region.

In this paper we present a dedicated search, in the four-top final state,for gauge singlet vector
bosons at the LHC at 14 TeV (LHC14). We focus on the same-sign
dilepton channel, where the Standard Model (SM) backgrounds are
small.\footnote{An early study of $t\bar{t}$ resonances in this
  channel~\cite{Lillie:2007hd} omitted an irreducible background
  which, although initially small, is a major component of the total
  background after cuts.}  We carefully
consider all leading SM background processes, simulating them using
merged and matched jets where necessary, and estimate the size of the leading
fake backgrounds.

The paper is organized as follows: in Section~\ref{sec:simplified}, we define a
simplified model for a Standard Model singlet vector boson $\rho$
coupling only to right-handed top quarks, and study its production and
decay at LHC14. We present the sensitivity
for discovery and exclusion in the parameter space of such a model in
Section~\ref{sec:results}, and give the 95\% exclusion limit on the
cross section$\times$branching ratio for a singly-produced
top-antitop resonance in the 4-top final state.  We interpret these
results in the context of a Composite Higgs scenario in
Section~\ref{sec:CHinterpretation}, and compare these to limits
obtained at 8 TeV in this and other channels. We summarize our results and conclude in
Section~\ref{sec:conclusion}.

\section{Massive singlet vector boson}
\label{sec:simplified}

We define a simplified model
with a canonically-normalized colour- and electroweak-singlet vector
boson coupled only to right-handed top quarks, as
follows:
\begin{equation}
\label{eq:simplifiedmodelrho}
\mathcal{L}_{\rho} = - \frac{1}{4} \rho_{\mu\nu}\,\rho^{\mu\nu}  +
\frac{1}{2}\, M_{\rho}^2 \,\rho^2+
g_{\rho}\,\bar{t}_R\, \slashed{\rho} \, t_R
\end{equation}
where being a singlet of the SM custodial symmetry , it can couple
neither to the transverse nor the longitudinal modes of the SM gauge bosons.
 For the latter, this is simply because one cannot obtain a spin-1 state with isospin-0 from two identitical isospin-triplet scalars due to Bose symmetry.

\begin{wrapfigure}{r}{0.5\textwidth}
\centering
\includegraphics[width=0.48\textwidth]{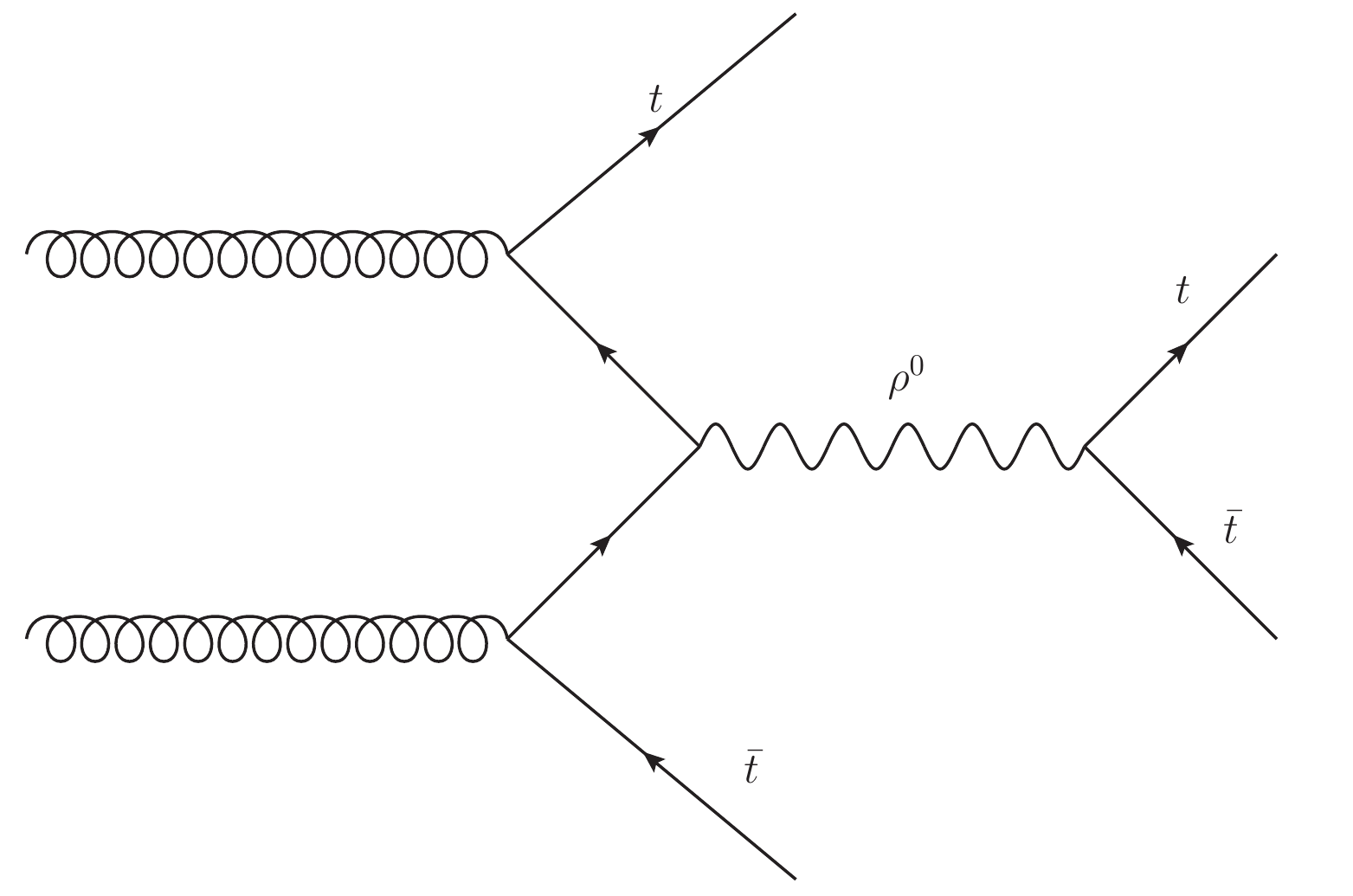}
\caption{Typical Feynman diagram for process $gg\rightarrow
  t\bar{t}\rho\rightarrow t\bar{t}t\bar{t}$.}
\label{fig:4top}
\end{wrapfigure} 
Both the production cross section and the decay width of these vector
singlets are controlled by their coupling to top quarks.  At typical LHC energies the top quark
content of the proton can be neglected, rather we consider gluons
in the initial state, splitting to high-$p_T$ top quark pairs. The
leading tree-level production process occurs via $t\bar{t}$
scattering, singly-producing the vector resonance in association with
a top-antitop pair.  The resonance subsequently decays to another
$t\bar{t}$ pair, resulting in a four-top final state (see Fig.~\ref{fig:4top}).

Production via a top loop, analogous to gluon-gluon fusion in higgs
production, is forbidden at leading order by the Landau-Yang
theorem~\cite{Landau:1948kw}.  The first non-zero contribution in the
$t\bar{t}$ final state must thus occur
at $\mathcal{O}(g_s^6 g_\rho^2)$, by emission of an additional
hard jet; this process is formally higher-order in $g_s$ than the
$t\bar{t}$ scattering process considered above, $\mathcal{O}(g_s^4 g_\rho^2)$, as well as suffering
from larger Standard Model backgrounds. 

There are also subleading effects that go in the opposite direction,
enhancing the relative sensitivity of the gluon-fusion process.
First, the cross section
  for the NLO top loop
  diagram will be enhanced by the valence quark component of the
  parton distribution function (PDF) in
  the initial state.  The gluon-initiated component will also be
  enhanced since it is evaluated at a smaller centre-of-mass
  energy (no production of additional top quarks).~\footnote{For a scalar
  resonance these effects enhance the top loop contribution by an
  order of magnitude over the naive expectation~\cite{Han:2014nja}.}. Finally, even
  though it suffers from a huge background from SM $t\bar{t}$, as
  mentioned above, its combinatorics are more tractable, allowing the
  resonance mass to be fully reconstructed in the semileptonic
  channel.  A definitive answer as to which process drives the
  sensitivity for $\rho$ would
  require computation of the loop and box diagram contributions
  to $\rho$ production, in the limit of small top mass. We consider
  this to be beyond the scope of the current analysis, and reserve it
  for future work~\cite{Liu:WIP}.  
For the remainder of this paper, however, we will assume that the
  naive power counting argument holds, and focus on the four-top final
  state.

Selecting the parameters $M_\rho=1$ TeV,
$g_\rho=1$ as a benchmark for illustrative purposes, the leading
order cross section is 4.88 fb, with a width-to-mass ratio for the $\rho$,
$\Gamma_\rho/M_\rho=3.6\%$.  The branching fractions for decays to 
the different final states are set by those of the $W$ boson, the
pure hadronic mode accounting for 31\% of the events; the single-,
di- and tri-lepton channels contributing 42\%, 21\% and 5\%
respectively, with the four-lepton channel contributing under
1\%\footnote{We have included leptonic tau decays in these counts.}.  

We plot the $p_T$ and $\eta$ distributions for
truth-level top quarks, ordered by
$p_T$, for $M_\rho=1$ TeV and $g_\rho$=1 in
Fig.~\ref{fig:parton} below.  One would nominally expect
to see two hard, central tops coming from the resonance decay, with $p_T\sim M_\rho/2$ (= 500
GeV for our benchmark), and two softer tops with $p_T\sim m_t=$173
GeV.  What we see instead is a rather more hierarchical spectrum after
$p_T$-ordering, implying a mixing between top quarks from different
origins.  In fact, although the leading top comes from the $\rho$ decay
almost 85\% of the time, if we ask that the two
hardest tops be daughters of the $\rho$, the probability falls to
50\%.  Note also that most of the top quarks are contained within the
central region of the detector $|\eta|<2.5$, as expected.  We also
plot the average number of top quarks per event with
$p_T>p_{T\text{min}}$ as a function of $p_{T\text{min}}$ for
$M_\rho=1,\,2$ TeV in Fig.~\ref{fig:parton}(c).  We see that for resonance masses
accessible at LHC14, we do not expect more than one top in each event
to be highly boosted ($p_T>1$ TeV).  The fully hadronic channel will thus contain a
large number of well-separated jets, the combinatorics making it very hard to
distinguish from QCD multijet background.  At the other extreme, the four-lepton
channel has too small a cross section.  In this work we focus on the
same-sign-dilepton channel, where we believe we will achieve the best
significance due to small SM backgrounds.
\begin{figure}[thb]
\centering
\subfloat[$p_{T}$ distribution for truth-level top quarks in signal
events with $M_\rho =$ 1 TeV and $g_\rho = 1$, where tops are $p_T$-ordered.]{%
\includegraphics[width=0.48\textwidth]{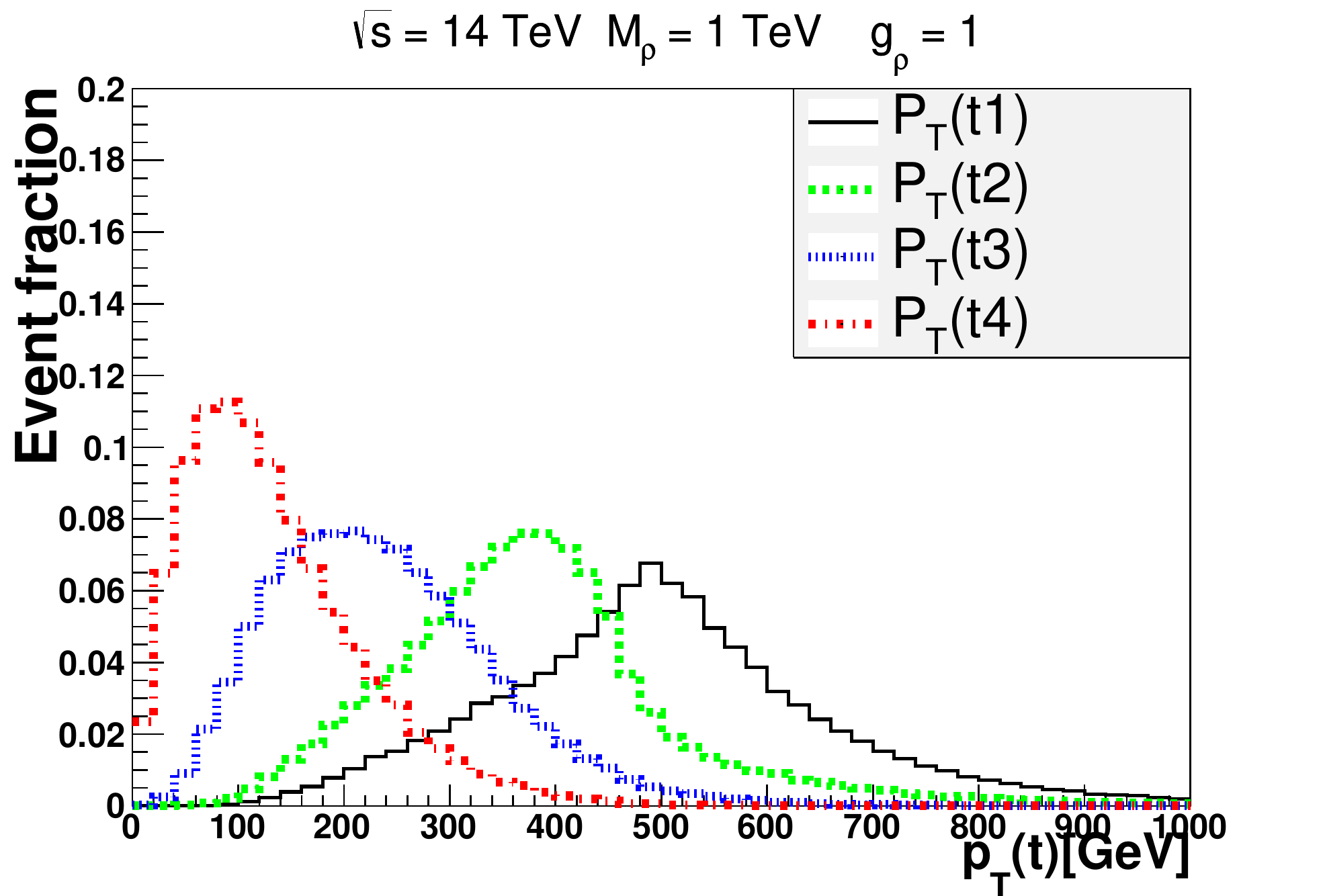}}
~
\subfloat[$\eta$ distribution for truth-level top quarks in signal
events with $M_\rho =$ 1 TeV and $g_\rho = 1$, where tops are $p_T$-ordered.]{%
\includegraphics[width=0.48\textwidth]{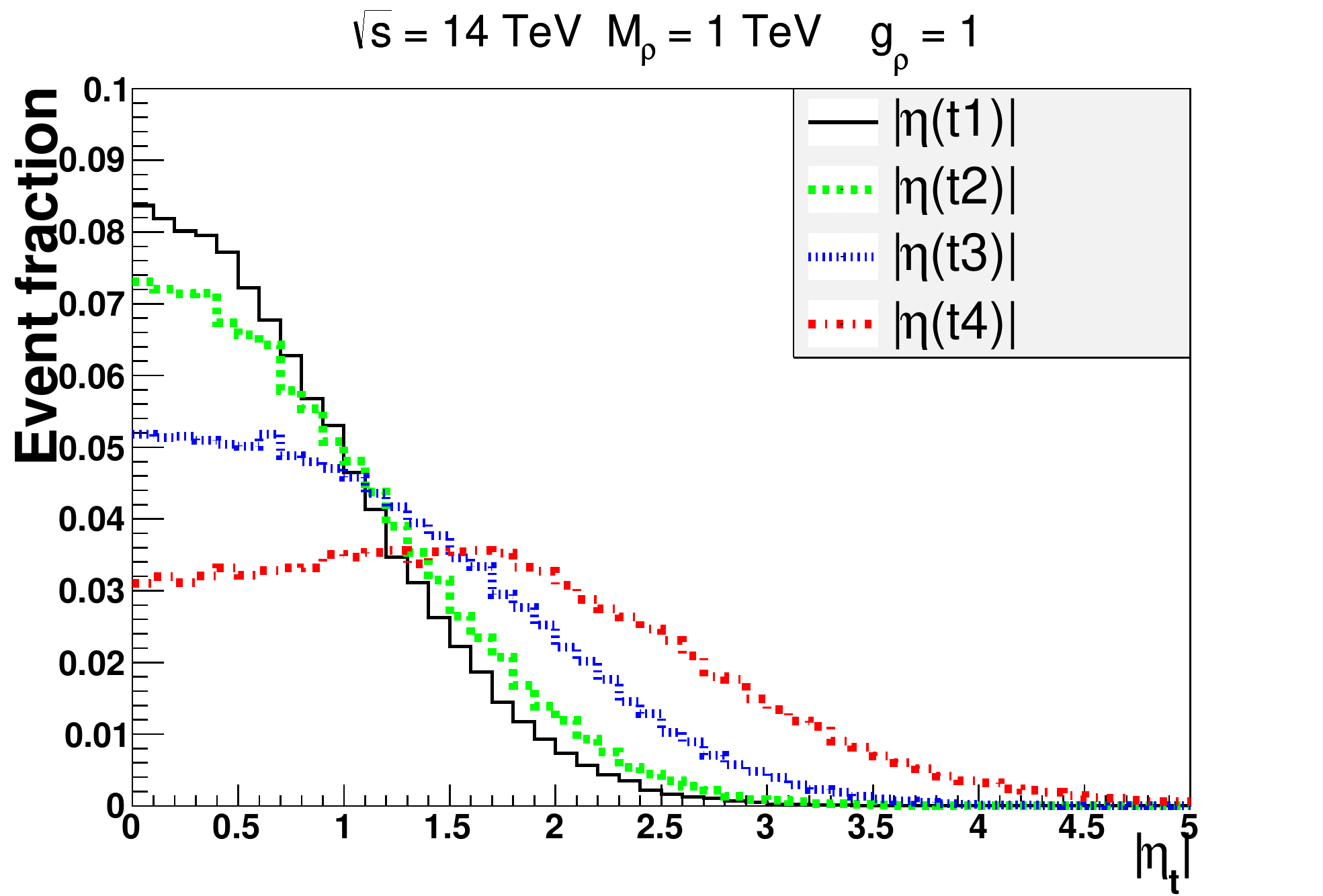}}
\\
\subfloat[The average number of truth-level tops per event with $p_T >
p_{T\text{min}}$ as function of $p_{T\text{min}}$ for $M_\rho = $ 1 TeV (black solid
) and  $M_\rho = $ 2 TeV (blue dashed).]{%
\includegraphics[width=0.48\textwidth]{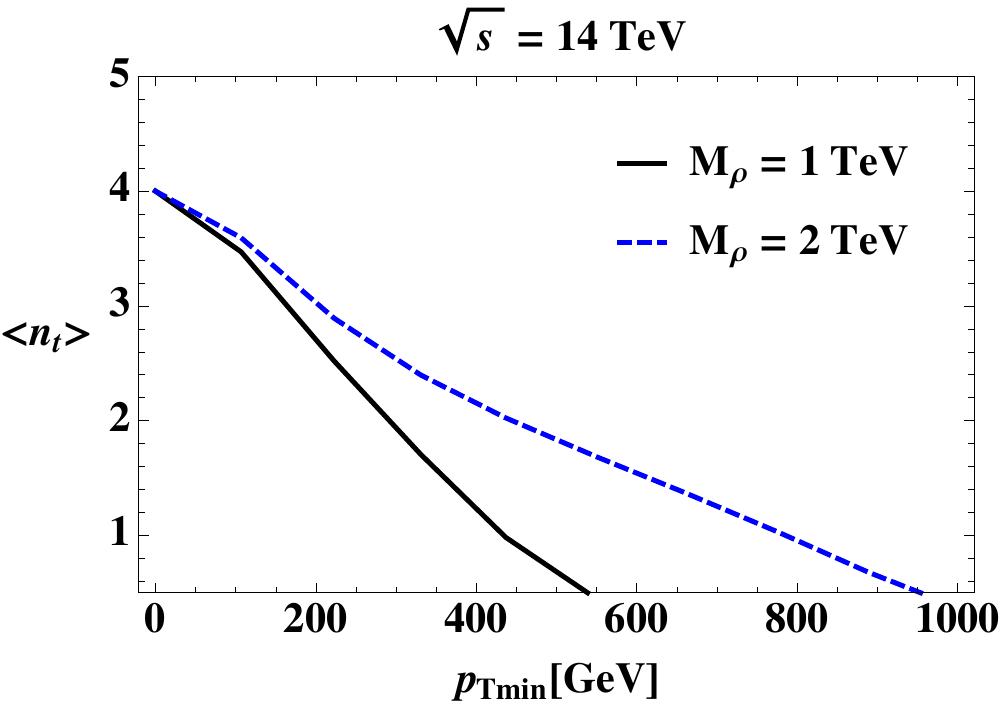}}
~
\subfloat[Normalized $\Delta R_{b\ell}$ distribution for truth-level $b$ quark and
lepton from daughter top quark, for $M_{\rho}
= 1$ TeV (black solid) and $M_{\rho} = 2$ TeV (red dotted).]{%
\includegraphics[width=0.48\textwidth]{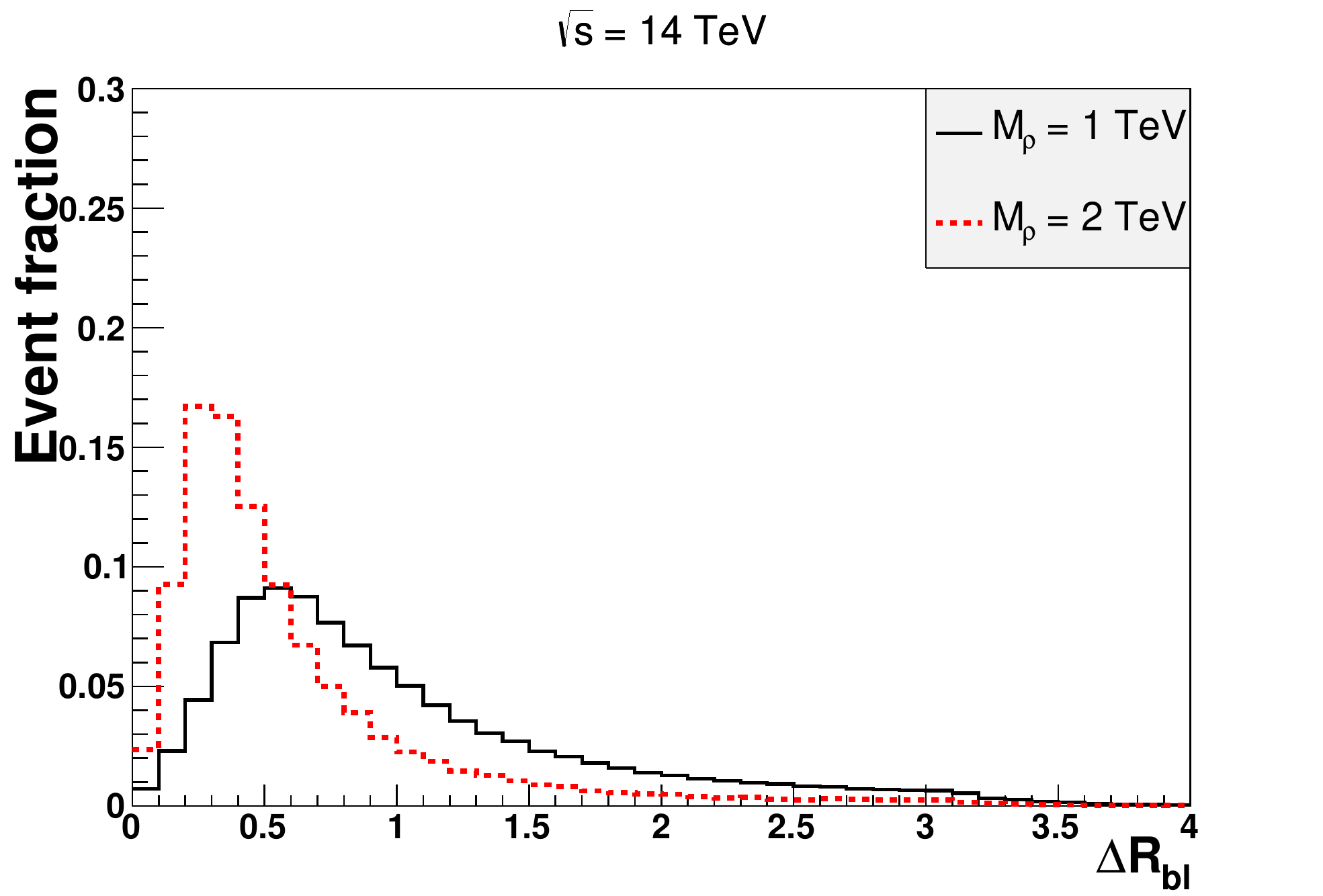}}
\caption{Truth-level distributions for top quarks and decay products in
  $pp\rightarrow t\bar{t}\rho\rightarrow t\bar{t}t\bar{t}$.}
\label{fig:parton} 
\end{figure} 

All results in this work were obtained by simulation using
\texttt{MadGraph5}~\cite{Alwall:2011uj}, interfaced to \texttt{Pythia
  6}~\cite{Sjostrand:2006za} for parton showering and hadronization as
needed. For the signal, we have implemented the simplified model using \texttt{FeynRules}~\cite{Feynrule} in UFO format. We use the CTEQ6L1 parton distribution function (PDF), in the
4-flavour scheme\footnote{This was shown in ~\cite{Maltoni:2012pa} to
  yield a good approximation to the result with large logs resummed at
  14 TeV.  Moreover this choice will only affect
    backgrounds that contribute under 2\% of the total, so any
    difference can be neglected.}
, and the default event-by-event renormalization and
factorization scales in \texttt{MadEvent}. \texttt{FastJet}~\cite{Cacciari:2011ma} was used to
reconstruct narrow jets, using the pre-implemented anti-kt algorithm
with $R=0.4$~\cite{Cacciari:2008gp}.  The signal was simulated at leading
order; backgrounds were simulated using matrix element-parton shower
merging and matching where necessary.  This was done using MLM matching, with $p_T$-ordered showers in Pythia, in the `shower-kT'
scheme, where the matching scale (QCUT = XQCUT) varied between 30
and 40 GeV, depending on the process.  The
cross-section of electroweak-boson-plus-jet
backgrounds were cross-checked using
\texttt{ALPGEN}~\cite{Mangano:2002ea}, interfaced to \texttt{Pythia 6}
for showering and hadronization.

With increasing $M_\rho$, we expect the leptons coming from top
decays will become increasingly collimated with the decay $b$-jet,
failing the standard fixed-cone isolation criterion (with $\Delta
R=0.3$) some
non-negligible fraction of the time.  This can be
clearly seen in Fig.~\ref{fig:parton}(d) above,
where we plot the normalized parton-level $\Delta R_{b\ell}$
distribution for leptonically-decaying $\bar{t}$ in the
signal, for two different resonance masses (1
and 2 TeV).  In order to retain as much
of the small signal cross section as possible, we use a modified
lepton isolation criterion.
\begin{wrapfigure}{r}{0.5\textwidth}
\centering
\includegraphics[width=0.48\textwidth]{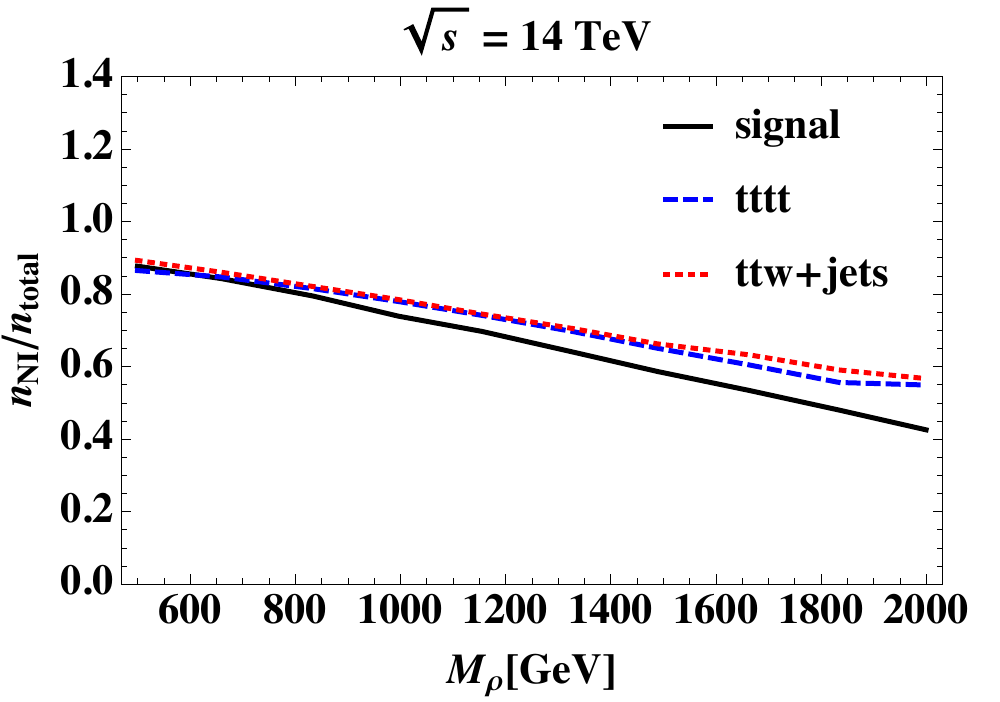}
\caption{Ratio of normal- and mini-isolation efficiencies for leptons, for signal (black solid) and two dominant
backgrounds: SM four-top (blue dashed) and $t\bar{t}W+$jets (red
dotted), after all cuts.}
\label{fig:miniISO}
\end{wrapfigure} 
  This was proposed
 by~\cite{Rehermann:2010vq}  as an efficient way to distinguish muons from top decays from those arising
from heavy flavour decays, and was subsequently successfully tested in Monte Carlo
studies of semileptonic top decays by ATLAS~\cite{Chapleau:2010nn}.
 The mini-isolation method involves applying an isolation criterion within a cone whose size
varies inversely with lepton $p_T$ (this quantity can be seen as a measure of
the boost of the parent) and requiring that the scalar sum of the hadronic
$p_T$ inside such a cone centred on the lepton be less than 10\% of
the lepton $p_T$.  Thus softer leptons are required to be more
isolated than harder ones. In Fig.~\ref{fig:miniISO}, we show the ratio of
efficiencies for lepton selection with regular and mini-isolation for the
signal and two dominant backgrounds.  Although the efficiency ratio is
similar for the signal and backgrounds, mini-isolation helps keep more
events after cuts, thus improving the significance over the entire
parameter space.  This improvement is especially important at high resonance mass, where
the production cross section is very small.

We define pre-selection cuts as follows:
\begin{eqnarray}
\label{eq:accept_cuts}
&p_{T,jcb} > 30 \,\text{GeV}, \qquad |\eta_j| < 4.5, \qquad |\eta_{cb}| < 2.5\\
&p_{T,\ell} > 25\, \text{GeV}, \qquad |\eta_{\ell}| < 2.5, \qquad
\sum\limits_{R_\text{min}} |p_{T,j}|\leq \; 0.1 \;p_{T,\ell}
\end{eqnarray}
where $p_{T}$ and $\eta$ denote the transverse momentum and
pseudorapidity of the reconstructed jets and mini-isolated leptons as
described above, and $R_\text{min}=\text{Min}(15/p_{T,\ell}\,,\,0.3)$.  A reconstructed jet is identified as a b(c)-jet if
its pseudorapidity satisfies $|\eta| < $ 2.5 and it is matched to a
b(c)-parton at  angular  distance $\Delta R < 0.2$.  We then require
exactly two same-sign leptons and at least 3 narrow jets.\footnote{We could in
principle exclude lepton pairs with an invariant mass inside the $Z$
mass window, to eliminate the contribution from $Z$+jets due to
charge-misidentification.  However, we estimate the contribution from
this subleading fake background to be negligible.}
In order to reduce the
backgrounds from di- and tri-boson plus jets, we stipulate at least 3
of the narrow jets be $b$-tagged.  We assume constant $b$-tagging and mistagging
efficiencies of 70\% for $b$-jets, 20\% for $c$-jets, and 1\% for light
jets, respectively.  We discuss the validity of this assumption in
Appendix~\ref{sec:btag}.  The $b$-tagging requirement ensures the dominance of
top-rich backgrounds, such as SM $t\bar{t}t\bar{t}$ and
$t\bar{t}W b\bar{b}$ production. There are
also large contributions from backgrounds with mis-tagged jets
    such as $t\bar{t}W$ + jets, as well as subleading contributions from
single-top in association with multiple vector bosons, where the
vector bosons decay to charm jets (35\% branching fraction for the $W$
boson).  A list of all leading
backgrounds with same-sign dileptons, including
their cross sections after pre-selection and cut efficiencies, is shown in
Table~\ref{tab:cutflow_s2}.  

\begin{figure}[htb]
\centering
\subfloat[Normalized distribution of number of b-jets, $n_b$ (defined in text) after preselection cuts.]{%
\includegraphics[width=0.48\textwidth]{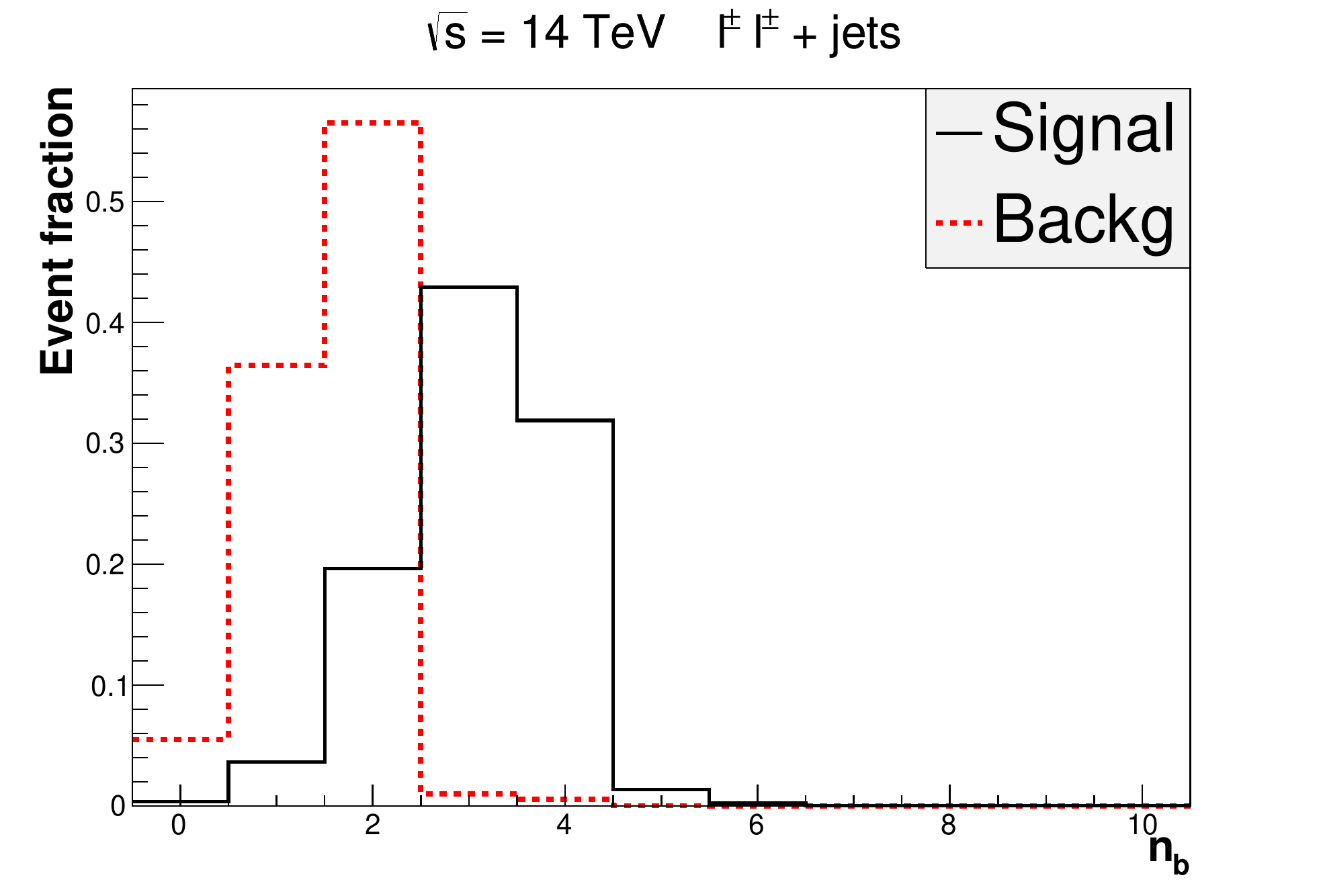}}
~
\subfloat[Normalized $H_T$ distribution for all reconstructed jets and leptons
after preselection and b-tagging.]{%
\includegraphics[width=0.48\textwidth]{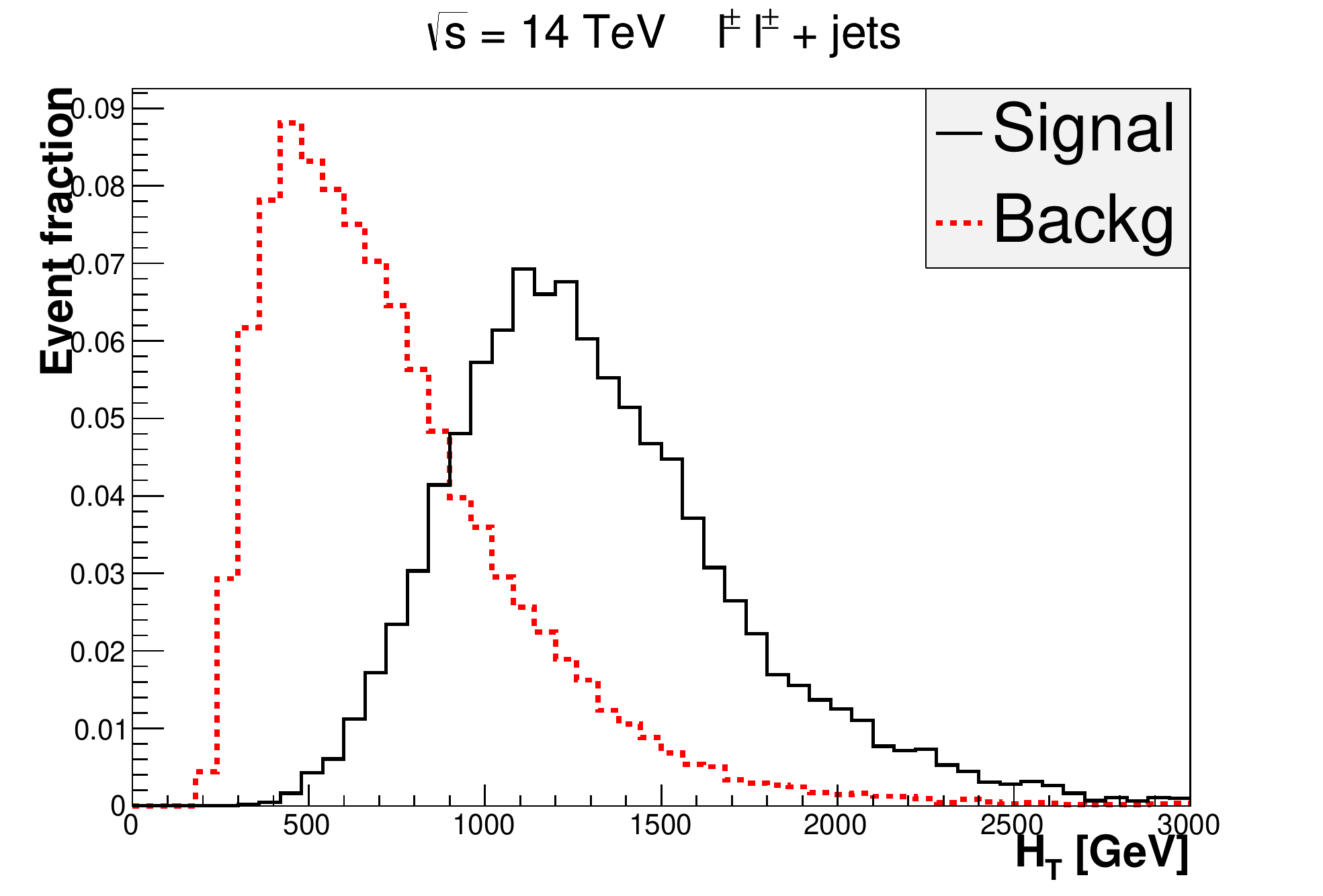}}
\caption{Comparing distributions for signal $pp\rightarrow t\bar{t}\rho\rightarrow
  t\bar{t}t\bar{t}$ with SM backgrounds containing same-sign
  dileptons, after preselection
  (for a full list see Table~\ref{tab:cutflow_s2}).}
\label{fig:htnj}
\end{figure}

We plot in Fig.~\ref{fig:htnj} the signal
and background distributions for the number of $b$-jets after preselection, and the
reconstructed $H_T$
distribution after requiring 3 $b$-tags, where $H_T$ is defined as the scalar sum of the $p_T$s of
the leptons and all reconstructed jets in the event.   This quantity
can be used as a proxy for the scale of the hard
scattering $\sim M_\rho + 2 m_t$, and as such, gives us some idea of
the mass of the resonance, which would be tricky to obtain by event
reconstruction due to combinatorics.  To further
suppress the backgrounds we put a hard cut on $H_T$, and require that
this be larger than the mass of the resonance (=1 TeV for our
benchmark model)
\begin{equation}
H_T=\sum\limits_{\text{all }j,\ell}\left|p_T\right| > M_\rho
\end{equation}
We verify that we have sufficient
statistics for all leading backgrounds, after all cuts have been
imposed.  We have not included K-factors in our results, since they are not
contained in the literature for many of our background processes.  We
expect the K-factor for our signal to be similar to that for SM
four-top production, which makes up a large component of the total
background.  We have also verified that changing the renormalization
and factorization scale to the more conventional $m_T/3$, where $m_T$
is the transverse mass of the $t\bar{t}\rho$ system, increases
the signal cross section by less than 20\%.

\begin{savenotes}
\begin{table}[thb]
\centering
\begin{tabular}{|c|c|c|c|c|c|}
\hline
\multirow{2}{*}{Process} & $\sigma_\text{pre}$ (ab)&
\multicolumn{2}{c|}{Cut efficiencies}&\multirow{2}{*}{$\sigma$ (ab)}\\
 &  SSDL + $n_j\ge$ 3    & $n_b\ge$ 3 & $H_T\ge 1$ TeV   &   \\
  \hline
\hline
  Signal ($M_\rho=$ 1 TeV; $g_\rho=1$)   &    161    & 0.43 & 0.78  &54.1  \\
\hline
  $t\bar{t}t\bar{t}$        &  224 & 0.39  & 0.37 &  31.9    \\
\hline
  $t\bar{t}W^\pm$+jets      & $8.43\times 10^3$& 0.026  & 0.16 &  34.2  \\
\hline
$t\bar{t} Z$\footnote{with one lepton from the $Z$ lost down the beampipe.} + jets  & $ 1.93 \times
10^3$& 0.024 & 0.14 & 6.71 \\
\hline
$t\bar{t} (h\rightarrow W^\pm W^{*\mp}\rightarrow \ell\nu qq)$  & $
1.21 \times 10^3$& 0.043 & 0.11 &  5.77\\
\hline
$t\bar{t} W^{+}W^{-}$ + jets   & 295 & 0.04 &  0.29 & 3.44   \\
\hline
$t \bar{t} W^\pm b\bar{b} $  & 21.6 & 0.31 & 0.22 & 1.50 \\
\hline
$t b W^+ W^- $  & 308 & 0.030 & 0.13 & 1.22 \\
\hline
$t b W^\pm Z $  & 155 & 0.029 & 0.15 & 0.661 \\
\hline
\hline
Total background        & & & &  85.4 \\

\hline
\end{tabular}
\caption{Cross sections for the signal and leading backgrounds
  containing same-sign dileptons (SSDL) after preselection, cut
  efficiencies for $b$-tagging
  and $H_T$ cut, for $M_\rho=1$ TeV and $g_\rho=1$. The last column shows the final cross sections after all the selection cuts. Leading backgrounds are
  merged and matched, including up to two extra jets where relevant.}
\label{tab:cutflow_s2}
\end{table}
\end{savenotes}

Since the number of signal event is very small after all the cuts, we must
also consider fake backgrounds, due to e.g. charge misidentification, or
jets faking leptons.  Contributing to the former will be $t\bar{t}+$j, and
  $Z+4b$; with semileptonic
$t\bar{t}$ and $Wj+4b$ for the latter.  We expect the $t\bar{t}$
background to be dominant in both instances, since it is produced at
lower order in QCD.  This expectation was confirmed in simulation,
yielding a cross section after cuts of 2.62$\times10^{3}$ ab in the
dileptonic channel, and 6.17$\times10^{4}$ ab in the semileptonic
channel.  We can make a crude estimate of the fake rate by applying a
constant efficiency for each, based on the CMS and ATLAS
TDRs~\cite{CMS:TDR,Aad:2009wy}. Using 10$^{-3}$ for charge mis-ID and
10$^{-5}$ for jets-faking-leptons, for example, yields a contribution
from fakes of less than 5\% of the total background cross section,
implying that these backgrounds are well under our control.  In
reality, however, the fake rates are strongly $p_T$-dependent, and a
detailed experimental study would be required to confirm our estimate.

\section{Results}
\label{sec:results}

\begin{figure}[htb]
\centering
\subfloat[Luminosity in fb$^{-1}$ required for discovery of spin-1
singlet resonance $\rho$ in SSDL channel of $4t$ final state, with
99.9999\% confidence (corresponds to $5\sigma$ in large-statistics limit).]{%
\includegraphics[width=0.48\textwidth]{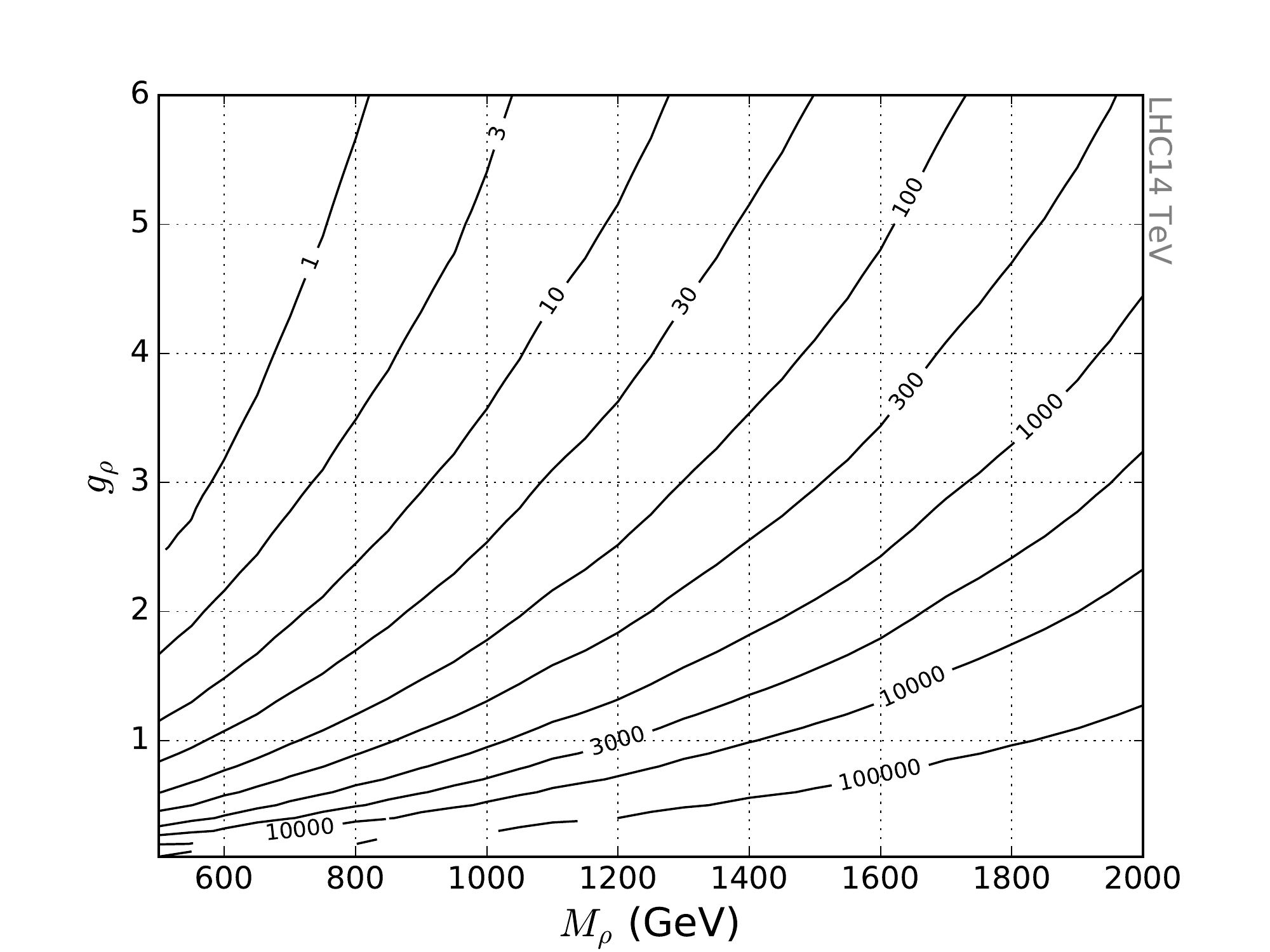}}
~
\subfloat[Expected exclusion for singly-produced $\rho$ in SSDL
channel of $4t$ final state, at 95\% confidence with SM
signal injection and luminosity of 30 (300) fb$^{-1}$.]{%
  \includegraphics[width=0.48\textwidth]{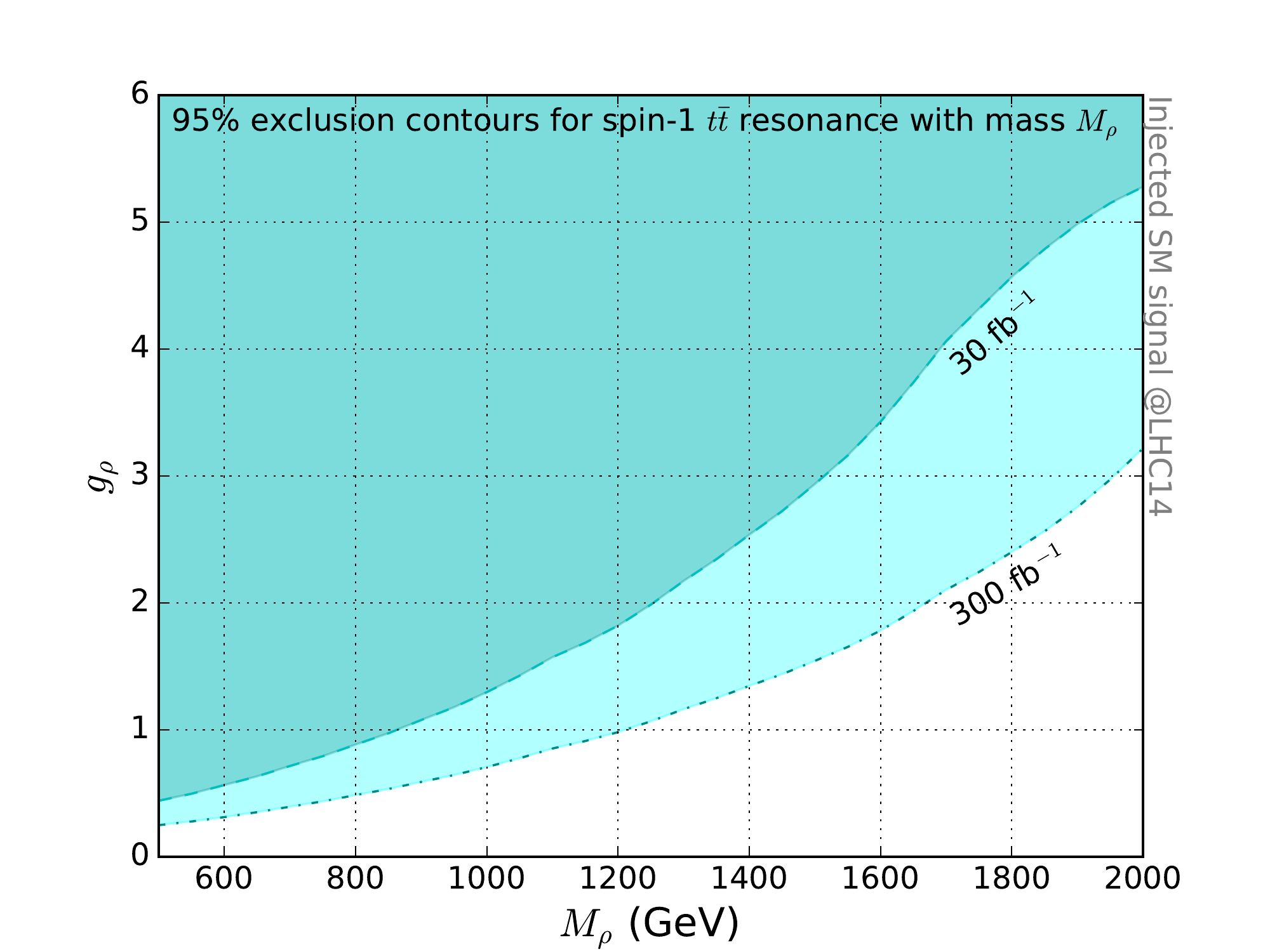}}\\
\subfloat[95\% expected upper limit on $\sigma\times BR$ for $t\bar{t}$ resonance produced in association with a
top-antitop pair with luminosity of 300 fb$^{-1}$ .]{%
\includegraphics[width=0.48\textwidth]{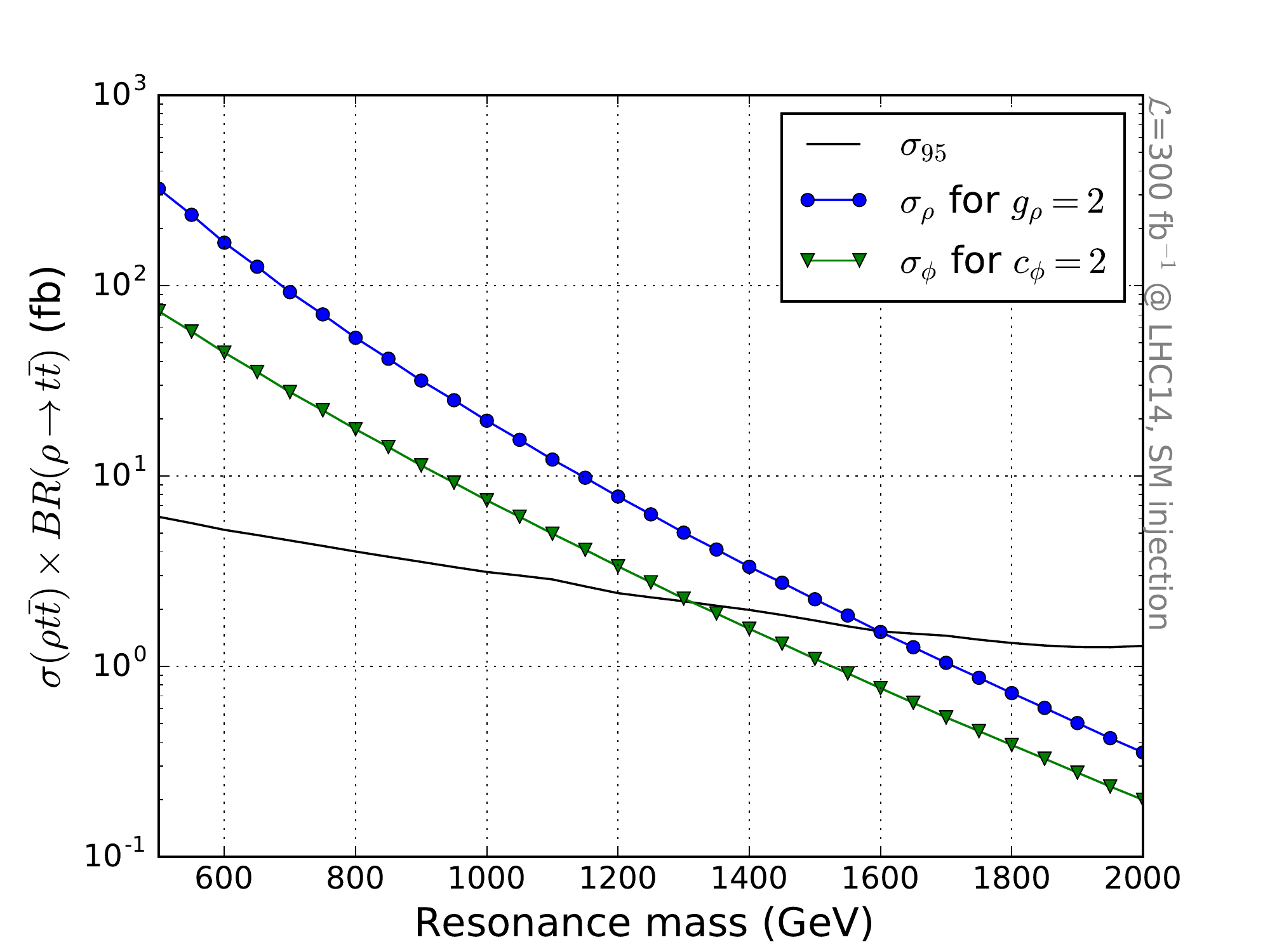}}
~
\subfloat[Luminosity in fb$^{-1}$ required for discovery of singlet
scalar $\phi$ in SSDL channel of $4t$ final state, with
99.9999\% confidence (corresponds to $5\sigma$ in large-statistics
limit).]{%
\includegraphics[width=0.48\textwidth]{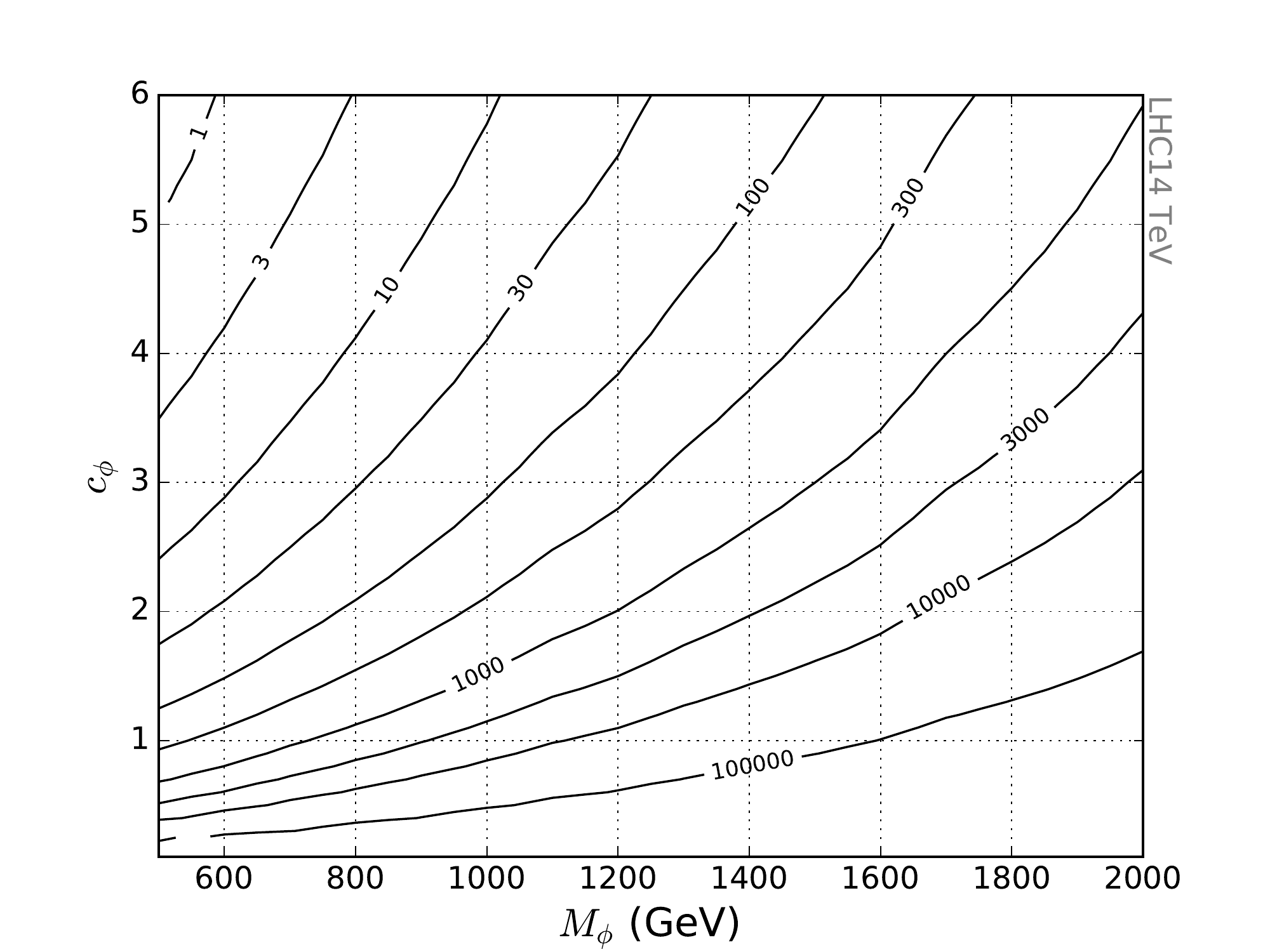}}
\caption{Discovery/exclusion potential for gauge singlet $t\bar{t}$
  resonance at LHC14 in $4t$ final state.}
\label{fig:results}
\end{figure} 

Our final results are shown in Fig.~\ref{fig:results}, with the
statistical procedure used to obtain them summarized in
Appendix~\ref{sec:stats}.  In Fig.~\ref{fig:results}(a), we
plot isocontours of the integrated luminosity required for discovery
of a gauge singlet spin-1 $t\bar{t}$ resonance at LHC14.  We naively
rescale the signal cross section computed for a coupling of unity
with $g_\rho$, in the narrow width approximation, ignoring
interference effects with SM 4-top production.  We justify this
simplification in Appendix~\ref{sec:finitewidth}. 
 
We see that at moderate (large) coupling,
$g_\rho=3\,(6)$, 300 fb$^{-1}$ of integrated luminosity at LHC14 will
allow us to discover a spin-1 singlet resonance up to $\sim$1.5 (1.9)
TeV.   Discovery of a resonance with smaller coupling, say $g_\rho$=2,
seems unlikely for masses larger than $\sim$ 1.3 TeV before the high-luminosity
upgrade of the LHC, although exclusion of
this region of parameter space should be possible with 95\%
probability by the end of LHC Run 3 (see
Fig.~\ref{fig:results}(b)).

Our results can also be used to compute the discovery reach/exclusion
potential in the $4t$ channel for any $t\bar{t}$ resonance that is
singly produced in association with a top-antitop pair, where the
kinematics (and hence the cut efficiencies) are likely to be similar
to those of the vector resonance.\footnote{This is not hard to imagine,
since our analysis is rather generic, and relies neither on
any sophisticated mass reconstructions, nor on particular spin-dependent effects.}  For ease of
recasting, we present our results as a 95\% exclusion limit on
$\sigma\times BR$ for this channel as a function of the resonance mass
with an integrated luminosity of 300 fb$^{-1}$
in Fig.~\ref{fig:results}(c).  

In particular we can trivially estimate
the discovery luminosity required for a spin-0 resonance $\phi$, with
a chiral-symmetry breaking coupling to top quarks of $c_\phi\,\phi\,\bar{t}_L\, \,t_R +\text{h.c.}$  Such a scalar
  could be found in a (fine-tuned) corner of the MSSM parameter space,
  for example, as the heavy higgs in the pseudoscalar
  decoupling limit, and for $\tan{\beta}\lesssim 3$.  Alternatively it
  could be the heavy pseudoscalar resonance in Superconformal Technicolor theories~\cite{Azatov:2011ps}.
The size of the
coupling $c_\phi$ will depend on the representation of $\phi$ under
the SM weak gauge group, $SU(2)_L$.  If it is a doublet, then $c_\phi$
can be $\mathcal{O}(1)$.  If it is an electroweak singlet, however,
then the above coupling is strongly suppressed, since it originates in a dimension-5 operator involving
the higgs field, with a coefficient $c_\phi=g_\phi m_t/\Lambda$, for
a cutoff $\Lambda$ that is
parametrically larger than the $\phi$ mass. The size of
  $g_\phi$ will depend on the origin of the interaction, for a
  weakly-coupled theory it must be of $\mathcal{O}(y_t)$, but it can be
  larger if it originates from a strongly-coupled sector.

Since the scalar couples to left-handed (LH) as well as RH top quarks,
we might expect the efficiency for lepton selection to change,
since leptons originating from decays of LH tops have smaller $p_T$,
due to preferential emission antiparallel to the parent top quark's boost.  However we
expect this to be a small effect, and hence apply the $\rho$
efficiencies naively.  We show the luminosity isocontours required for
discovery of a scalar resonance in Fig.~\ref{fig:results}(d).  As
expected, the results for a scalar resonance are not quite as
encouraging as those for the vector resonance, particularly if the
scalar is a gauge-singlet elementary field, in which case $c_\phi$ is
constrained to be rather small.  Instead, we expect the sensitivity
for the scalar resonance to be driven by the $t\bar{t}$ final state,
since the gluon-fusion production is unsuppressed, and rather large.

In principle it should be possible to compare the sensitivity of our
analysis to that of other searches for $t\bar{t}$ resonances.  One
example is the 8 TeV ATLAS resonance search in the lepton-plus-jets channel of
the $4t$ final state~\cite{Aad:2015kqa}.  Their results are presented in
the form of exclusion limits on $\sigma\times BR$, but here the
benchmark resonances used to obtain these results are pair-produced,
resulting in a much larger $H_T$ in the final state than in the case
of single production, for a resonance with equal mass.
This would give rise to large differences in the efficiencies for
their $H_T$ cuts, and we cannot simply recast their limits in the
context of our simplified model.  

ATLAS also present their results as
limits on the coupling of a four-top contact interaction of the
following form:
\begin{equation}
\frac{C_{4t}}{\Lambda^2}\left(\bar{t}_R\gamma^\mu
  t_R\right)\left(\bar{t}_R\gamma_\mu t_R\right)
\end{equation}
which might also be useful for the purposes of comparison.
Using a likelihood fit to the $H_T$
spectrum after cuts to LHC data at 8 TeV centre-of-mass energy, they
obtain a 95\% CL upper limit on the coefficient of the 4$t$ contact
interaction $|C_{4t}|/\Lambda^2 < 6.6$ TeV$^{-2}$. By integrating out
the $t\bar{t}$
resonance, we can naively interpret this as a
limit on the relevant combination of our
simplified model parameters, yielding $M_\rho/g_\rho > 275$ GeV. However, care
must be taken to ensure that this limit is consistent with the
effective theory being used within its regime of validity in the
analysis.  In this particular instance the limit is obtained by a
comparison of their measured $H_T$ distribution to that expected from
signals and backgrounds, over the entire range of $H_T$ measured ($\sim$
2 TeV).  In the absence of any information to the contrary, we have to
assume that the entire
range of $H_T$ was equally instrumental in deriving the limit, and
since $H_T$ can be thought of as a lower bound for the centre-of-mass
energy, their limit can only be applied for $M_\rho> 2$ TeV.  Hence
their limit cannot be applied for $g_\rho\lesssim 7$!

When set in the
  broader context of a realistic scenario, there will also be additional
  constraints on singlet bosons due to their subleading interactions.  We will explore some of these in
  the context of the $SO(5)/SO(4)$ composite higgs in
  Section~\ref{sec:CHinterpretation} below.

\section{Interpretation in Composite Higgs framework}
\label{sec:CHinterpretation}

The encouraging results obtained in the large-coupling region of our
simplified models beg for an interpretation
within the Composite Higgs (CH) framework, in which the Higgs arises as a
pseudo-goldstone boson of some larger
global symmetry
(see~\cite{Contino:2010rs,Panico:2015jxa} for a comprehensive review, and references therein).
The presence of spin-1 resonances is a robust prediction
  in this framework, as they can be excited from the vaccum by the
  conserved currents in the strong sector. In typical CH models, however, it is the composite fermion resonances that are usually
assumed to be among the lightest new states in the theory, since these
are expected to cut off the large top-quark loop contribution to the
quadratic divergence of the higgs mass.  Furthermore, there are
usually strong constraints on the mass of vector resonances that are
electroweak- or colour-charged, from precision electroweak
measurements, and flavour-changing neutral currents, respectively.
These stringent limits do not apply to singlet resonances however,
hence there is no theoretical bias against a composite vector resonance being the
lightest new particle in the theory, provided it is a gauge singlet.

A gauge-singlet spin-1 resonance   is, in fact, present in
many simple incarnations of this scenario, excited by the conserved
current of a global $U(1)_X$ symmetry group. Such a group is required
in order to 
correctly reproduce the hypercharge of the RH top quark, in (more minimal) scenarios where the $t_R$ is a composite singlet of the
strong-sector global symmetries. This resonance, which we denote as $\rho_X$, only interacts with elementary fermions through
small mixing terms, suppressed by powers of the ratio
$g^\prime/g_{\rho_X}$, where $g^\prime$ is the coupling of the SM hypercharge
gauge boson (which mediates the coupling of $\rho_X$ with the rest of
the elementary sector via a linear mixing), and
$g_{\rho_X}$ is a large coupling typical of the composite sector.  Among the SM fermions, the right-handed top alone is not constrained
to be a purely elementary field; in the case that it is a fully
composite singlet under the global symmetries, it could have
a large coupling to $\rho_X$, as in the CH model with a minimal $SO(5)/SO(4)$
coset structure:\footnote{ We assume a large separation of
scales between
the mass of the singlet bosons and all heavier mass scales in the
theory, including other composite states, and integrate out the
latter.}
\begin{equation}
\mathcal{L}_{\rho_X}=- \frac{1}{4} \rho_{\mu\nu}^X \rho^{X \mu\nu} + \frac{m_{\rho_X}^2}{2
  g_{\rho_X}^2} (g_{\rho_X} \rho^X_\mu - g_{el}^\prime B_\mu)^2\, + c\,
\bar{t}_R\gamma^\mu(g_{\rho_X} \rho^X_\mu - g_{el}^\prime B_\mu)
t_R + \cdots
\end{equation}
Here $c$ is an $\mathcal{O}(1)$ parameter which we set equal to 1
for simplicity, and we are omitting additional higher derivative interactions that stem from the CCWZ
construction.\footnote{We also treat the mass
and coupling as independent parameters, although in the
SILH~\cite{Giudice:2007fh} power-counting, they are related, via the
global-symmetry-breaking scale $f$, to a measure of the fine-tuning in
the higgs mass.}  The full lagrangian and interactions can be found in
~\cite{Greco:2014aza}, with important intermediate results summarized in
Appendix~\ref{sec:so5modso4} for convenience.

\begin{figure}[htb]
\centering
\subfloat[Cross section contours (fb) for on-shell production of the
$\rho_X$   The black solid (blue dashed) line corresponds to the
production via $t\bar{t}$ fusion (Drell-Yan type production).  The
$t\bar{t}$ fusion cross section dominates
in the shaded region.]{%
\includegraphics[width=0.48\textwidth]{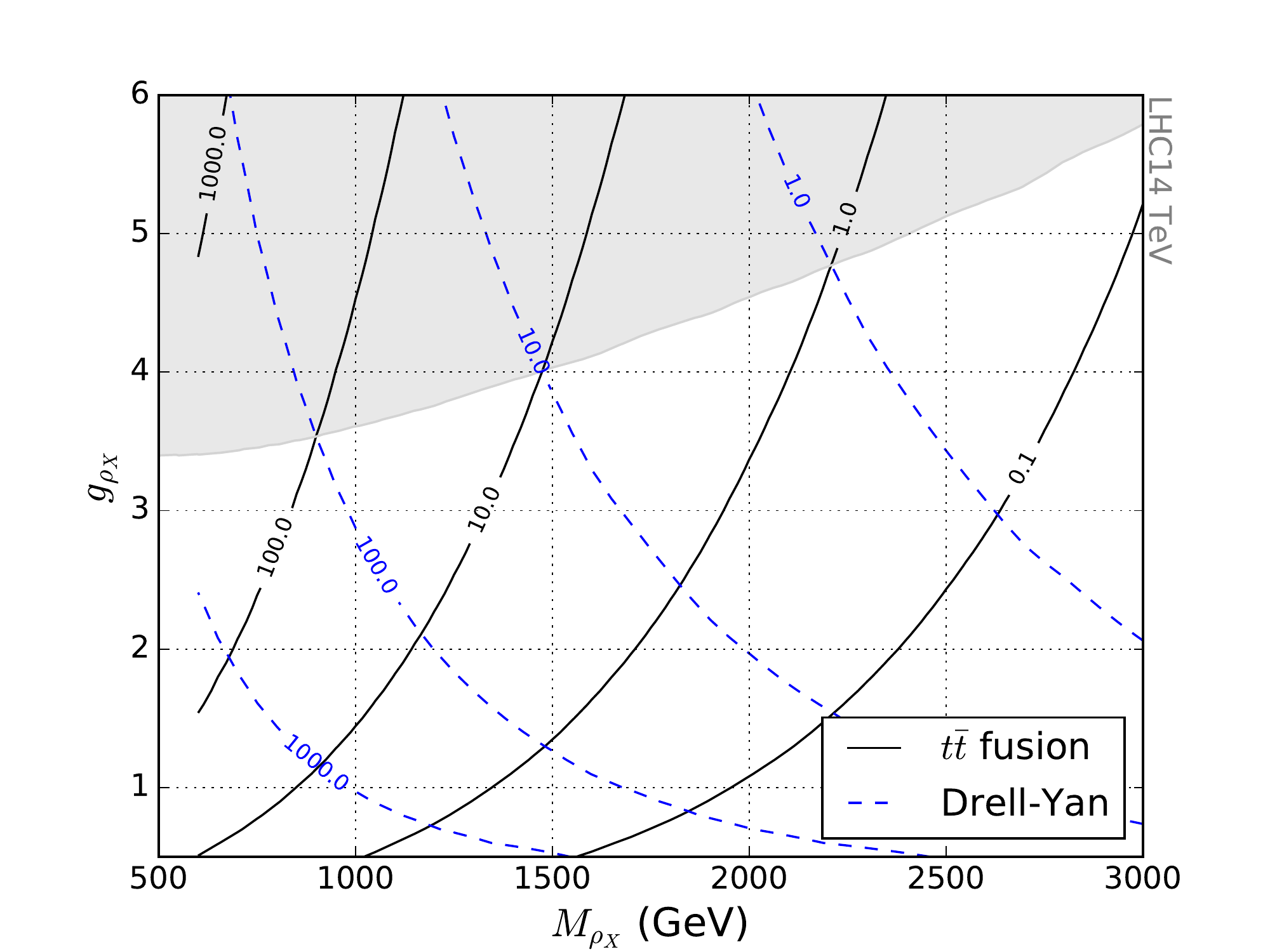}}
~
\subfloat[$\rho_X$ branching fraction as a function of coupling
$g_{\rho_X}$ (mass-independent for large $M_{\rho_X}$).]{%
\includegraphics[width=0.48\textwidth]{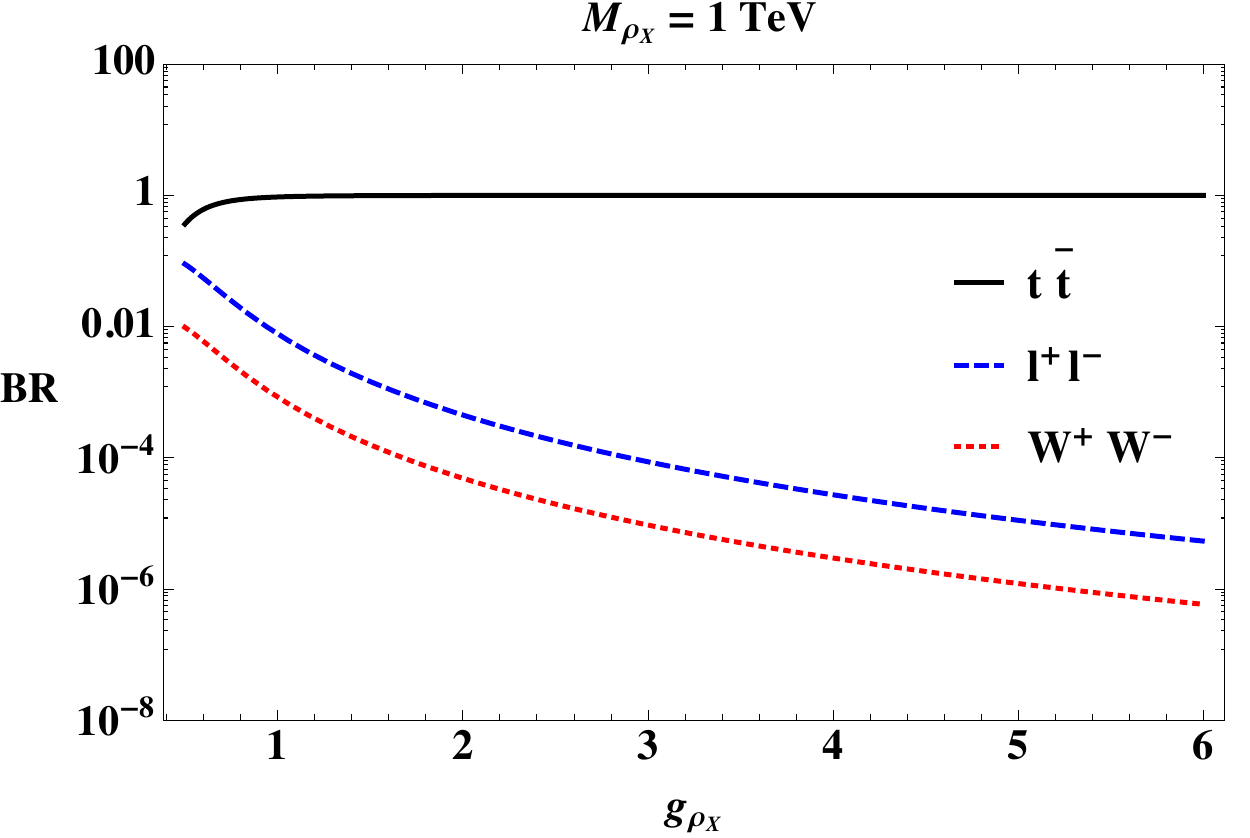}}
\caption{Production cross section [fb] and decay branching fraction for spin-1 singlet
  boson $\rho_X$, with coupling $g_{\rho_X}$ to RH tops, and a mixing-suppressed
coupling $g^\prime/g_{\rho_X}$ to elementary fermions.}
\label{fig:xs}
\end{figure}

\begin{wrapfigure}{r}{0.5\textwidth}
\vspace{-0.3in}
\centering
\includegraphics[width=0.48\textwidth]{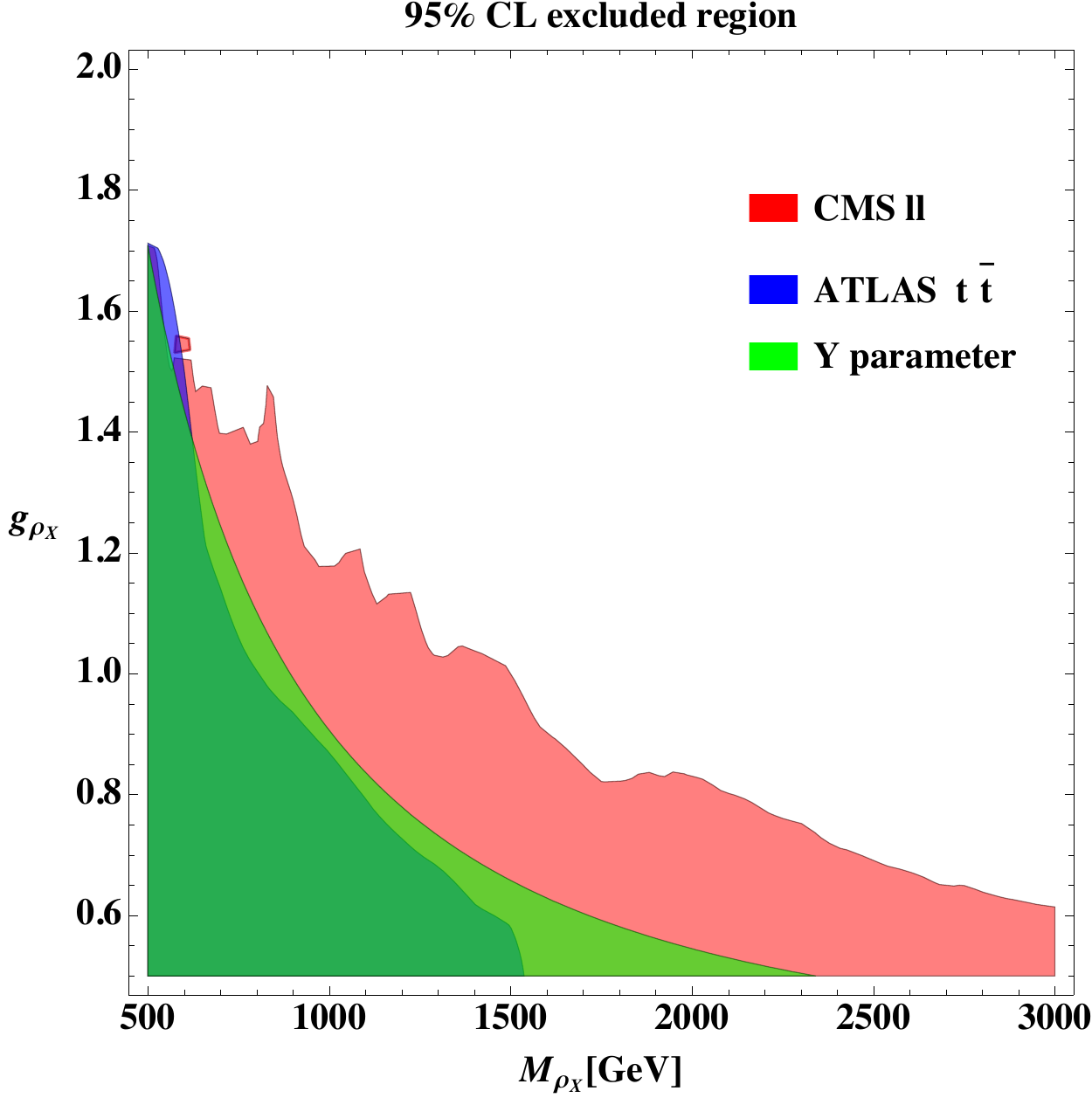}
\caption{Current limits on the $\rho_X$ parameter space in CH models
  from most sensitive 8 TeV analyses:  CMS dilepton
  search~\cite{Khachatryan:2014fba} and ATLAS
  $t\bar{t}$~\cite{Aad:2015fna}. We also include the constraint
  from the electroweak $Y$ parameter (details in text).} 
\label{fig:massbound}
\end{wrapfigure}

As mentioned above, through linear mixing with the SM hypercharge 
gauge boson, $\rho_X$ will also acquire (mixing-suppressed) couplings
to other SM states, such as $W$ bosons and elementary quarks\footnote{Because of its singlet nature, the couplings with SM  $W$ gauge boson can only arise after EWSB.}.  These
give rise to additional production mechanisms for $\rho_X$, via
vector-boson fusion (VBF), or a Drell-Yan-like process
$q\bar{q}\rightarrow \rho_X\rightarrow t\bar{t}$, as well as
additional decay modes.  Drell-Yan production is suppressed 
with respect to production via $t\bar{t}$ fusion considered
above, by a factor of $g^{\prime 4}/(g_{\rho_X}g_s)^4$ ;
VBF is further
suppressed by the $W$ PDF inside the proton, and is effectively
negligible~\cite{Greco:2014aza,Pappadopulo:2014qza}. In the large mass region, however,
the top fusion channel falls much faster than the Drell-Yan
contribution, due to the steep drop of the gluon PDFs at large $x$.  We show the competing effect of $\rho_X$ production in
the two leading channels, as well as its decay branching fractions for
fixed mass (branching fractions are almost independent of mass in the
large $M_\rho$ limit) in Fig.~\ref{fig:xs}.
In the small-coupling region, it
may be more effective to search for the $\rho_X$ boson in one of its
alternative decay modes, via Drell-Yan type production.  
Various
searches in relevant channels have been carried out by the ATLAS and
CMS experiments, with results presented in terms of limits on
$\sigma \times BR$  for each channel.  The search with the
largest sensitivity over the entire range of $\rho_X$ masses considered
in this work are the ATLAS and CMS high-mass dilepton resonance
searches~\cite{Aad:2014cka,Khachatryan:2014fba}.  Since
$\sigma\times BR$ in this channel scales like
$(g^{\prime 4}/g_{\rho_X}^2)\times(g^{\prime}/g_{\rho_X})^4$, however, the
limit becomes quickly irrelevant above $g_{\rho_X}\sim 1.2$, where the
ATLAS $t\bar{t}$~\cite{Aad:2015fna} search takes over in sensitivity, the branching ratio
to $t\bar{t}$ exceeding 90\% above $g_{\rho_X}=1$ (see
Fig.~\ref{fig:xs}(b)).  Other searches, e.g. in the
$WW$\cite{Aad:2015ufa, Aad:2015owa, Khachatryan:2014gha,Khachatryan:2014hpa},
$ZH$~\cite{Aad:2015yza,Khachatryan:2015bma,Khachatryan:2015ywa} and
$\tau\tau$ channels~\cite{Aad:2015osa}, as well as searches in dijets~\cite{Khachatryan:2015sja}, have negligible sensitivity and are not
considered here.
Fig.~\ref{fig:massbound} below we show the exclusion limits
on the $\rho_X$ parameter space recast from the two most sensitive
analyses, the ATLAS $t\bar{t}$~\cite{Aad:2015fna} and CMS dilepton
  searches~\cite{Khachatryan:2014fba}.
We see that the strategy advocated in this paper is exactly
complementary to existing searches in other channels, giving an
enhanced sensitivity at large $g_{\rho_X}$, which is not accessible by
other means.  Note that only the Drell-Yan-type production cross
section was used to set the limit in the $t\bar{t}$ channel.  In
principle there will also be a contribution due to gluon-gluon fusion
at next-to-leading order, but we expect this to be negligible in the
range of $g_{\rho_X}$ constrained here.  
Care must be taken in translating these limits
  on $g_{\rho_X}$ to limits on the simplified model parameter $g_\rho$,
  which are related as detailed in Appendix~\ref{sec:so5modso4}.
  Their difference is negligible in the limit of large $g_{\rho_X}$, but
  could be significant, and model dependent, for small values. 

There are additional constraints on the mass and coupling of $\rho_X$,
coming from precision electroweak observables.  The $Y$-parameter,
the 2nd derivative of the hypercharge form
factor~\cite{Barbieri:2004qk}, is the leading constraint here, since
there is no contribution to the $S$ parameter from a singlet.  To
compute the contribution to this low-energy observable from $\rho_X$ we simply integrate it out by
setting it equal to its equation of motion, giving at leading order in
the derivative expansion, the following terms in the effective lagrangian
\begin{equation}
\mathcal{L}_\text{eff} \supset
-\frac{1}{2}\frac{g_{\rho_X}^2}{m_{\rho_X}^2}\bar{t}_R\gamma^\mu
t_R\bar{t}_R\gamma_\mu
t_R-\frac{1}{2}\frac{g^{\prime 2}}{g_{\rho_X}^2m_{\rho_X}^2}\partial^\mu
B_{\mu\nu}\partial_\alpha B^{\alpha\nu}
\end{equation}
The second term yields an expression for $Y$ at tree-level, which can be constrained
using the global fit in~\cite{Barbieri:2004qk}\footnote{we ignore loop-
  suppressed contributions for simplicity}:
\begin{equation}
|Y|=\frac{g^{\prime 2}m_W^2}{g_{\rho_X}^2m_{\rho_X}^2}<1.2\times
10^{-3}\Rightarrow g_{\rho_X}m_{\rho_X}\ge 836\;\text{GeV}
\end{equation}
This is a rather weak limit; $Y$ is usually suppressed with
respect to the $S$-parameter by a factor of $g^{\prime 2}/g_{\rho_X}^2$. We see in Fig.~\ref{fig:massbound}(b) that this constraint is comparable
to that from the ATLAS $t\bar{t}$ search, which is, itself, not very
constraining for large values of $g_{\rho_X}$.  It is easy to see in
this plot the
complementarity between the sensitivity of current
search strategies, and the strategy we advocate in this paper.  It is
clear that $t\bar{t}$ fusion drives the sensitivity at larger
  couplings.


\section{Conclusion}
\label{sec:conclusion}
In this paper, we studied the reach for a top-antitop
vector resonance in the same-sign dilepton channel of the 4-top final
state at LHC14.  For a vector resonance that couples dominantly to top quarks, this
$t\bar{t}$ fusion channel is the leading tree-level production mode;
single production via a top loop being forbidden by Yang's
theorem. Our analysis made use of the large $b$-jet multiplicity of
the signal, as compared with the background, as well as the relative
paucity of Standard Model backgrounds with same-sign dileptons.  Due
to the large combinatorics of the 4-top final state, we did not attempt
a full reconstruction of the event, placing instead, a hard $H_T$ cut
on the reconstructed objects in the final state in order to select
events with higher centre-of-mass energies.  We found that the
irreducible SM
4-top background, which was omitted in a similar search, was a dominant
component of the background after cuts.  

We presented our results in the
form of isocontours of luminosity required for discovery, in the
parameter space (mass, coupling) of the
resonance, as well as a 95\% exclusion limit on the
cross-section $\times$ branching ratio in this final state (see
Figs.~\ref{fig:results}).  We found
a discovery reach (95\% exclusion) for vector resonances with 300 fb$^{-1}$ integrated
luminosity, of mass up to 1.2 (1.6) TeV for a coupling to right-handed
tops, $g_\rho=$2.  We also placed limits on a scalar $t\bar{t}$
resonance, although we expect the sensitivity in this case
will be larger in the $t\bar{t}$ final state. 

We interpreted our results within Composite Higgs scenarios, many
simple implementations of which contain a singlet vector resonance $\rho_X$,
excited from the vacuum by the conserved current of a $U(1)_X$ global
symmetry.  These vector singlets can
have a large coupling to RH top quarks in the case where the latter
are composite singlets of the strong sector. However they only
interact with other SM particles via a linear mixing with $B^\mu$,
the hypercharge boson, resulting in couplings that scale parametrically as
$g^\prime/g_{\rho_X}$.  Hence direct searches for these resonances decaying
to pairs of Higgs/gauge bosons, leptons, or light jets, have maximum
sensitivity for small $g_{\rho_X}$.  The most efficient way to
access the region of large $g_{\rho_X}$ is likely through the four-top final
state. Unfortunately existing resonance searches in the four-top channel
are not directly applicable to this class of
models, since their results are expressed either in terms of benchmarks with {\it
  pair-produced} resonances, or limits on the coefficient of a
four-top contact interaction.  For a light resonance that is
singly-produced via $t\bar{t}$ fusion, neither one applies. Its cut
efficiencies, particularly for hard $H_T$ cuts, are likely to be
considerably smaller than the corresponding ones for a pair-produced resonance of
the same mass. Moreover, the analysis appears to obtain much of its
sensitivity from events with a large centre-of-mass energy (up to
$H_T=2$ TeV), and hence
cannot be used to place limits on a four-top contact interaction
obtained by integrating out a resonance with mass smaller than this
scale.  For these reasons, we strongly urge the relevant experimental
groups to include in their benchmarks an example of a resonance that
is singly-produced, in association with tops, in order to improve
their coverage of the available theory space in this rather
well-motivated scenario.

\section*{Acknowledgements}
The authors wish to thank Roberto Contino, Riccardo Rattazzi,
Francesco Riva and Minho Son for many helpful discussions, and the CERN
theory group for its warm hospitality.  RM's work
was funded by the Swiss National Science Foundation under grant number
CRSII2 141847, ``Particle physics with high-quality data from the CERN LHC''.


\appendix


\section{Cross section tables}
\label{sec:tables}

In this appendix, we present the cross sections under different mass
hypotheses for the spin-1 (Table~\ref{tab:sigrho}) and scalar
(Table~\ref{tab:sigphi}) resonances, for production  through
$t\bar{t}$ fusion with unit coupling $g_\rho = c_\phi =1$.  These
cross sections were used in our determination of the 95\% upper limit
for the cross section. The cross sections were calculated using the
MadGraph5~\cite{Alwall:2011uj}, using the default event-by-event factorization and renormalization scales.  We also show the final cross sections for the signal and the total backgrounds after all the cuts for the spin-1 resonance $\rho$. In addition, we  present  in Table~\ref{tab:sigrho} the naive significance, $S/\sqrt{B}$, for the integrated luminosity of 300 fb$^{-1}$.
\begin{table}[ht]
\begin{center}
\begin{tabular}{|c|c|c|c|c|c|c|c|c|c|c|c|c|c|c|c|c|c|c|c|}
\hline
  $M_\rho$ [GeV] &  500 & 600  & 700 & 800 & 900  &  1000 & 1100  & 1200   \\
  \hline
\hline
$ \sigma_\rho$[fb] & 80.6 & 42.0  & 23.1 & 13.3 & 7.93 & 4.88  & 3.05 & 1.95  \\
\hline
  $\sigma_S$[ab] & 854 &470 & 262 &  151 & 89.4 & 54.1 &32.3 & 21.0   \\
\hline
  $\sigma_B$[ab] & 309 & 250  & 197 &  151 & 114 & 85.4 & 64.0  & 47.1 \\
\hline
$S/\sqrt{B}  $ &27   &  16 &10 & 6.8 & 4.6 & 3.2   &  2.2 &1.7 \\
\hline
\hline

  $M_\rho$ [GeV] & 1300 & 1400 & 1500 & 1600 & 1700 & 1800 & 1900 & 2000   \\
\hline
\hline
 $ \sigma_\rho$[fb] & 1.26 & 0.834 & 0.562 & 0.379 & 0.261 & 0.181 & 0.126 & 0.0883     \\
 \hline
  $\sigma_S$[ab]  &  12.8& 8.22 &  5.40 & 3.53 & 2.22 & 1.52 & 1.02 & 0.668 \\
\hline
  $\sigma_B$[ab]  &  34.0 & 24.7 &  18.0 & 13.4 & 10.1 & 7.82 & 5.98 & 4.61 \\
\hline
$S/\sqrt{B}  $& 1.2 & 0.91 & 0.70  & 0.53 & 0.38 & 0.30 & 0.23 & 0.17 \\
\hline
\end{tabular}
\end{center}
\caption{Production cross section, $\sigma_\rho$ for spin-1 resonance of mass
  $M_\rho$ for fixed coupling to the right-handed top quark $g_\rho$ = 1.
  Also shown is cross section after cuts ($\sigma_S$), background
  cross section ($\sigma_B$), and naive significance, $S/\sqrt{B}$, for integrated
    luminosity of 300 fb$^{-1}$.}
\label{tab:sigrho}
\end{table}
 
\begin{table}[ht]
\begin{center}
\begin{tabular}{|c|c|c|c|c|c|c|c|c|c|c|c|c|c|c|c|c|c|c|c|}
\hline
  $M_\phi$ [GeV] &  500 & 600  & 700 & 800 & 900  &  1000 & 1100  & 1200   \\
  \hline
\hline
$\sigma_\phi$ [fb] &18.3 & 11.1 & 6.9 & 4.4 & 2.8 & 1.9 & 1.2 & 0.84\\
\hline
\hline
  $M_\phi$ [GeV] & 1300 & 1400 & 1500 & 1600 & 1700 & 1800 & 1900 & 2000   \\
\hline
\hline
$\sigma_\phi$ [fb] & 0.57 & 0.40 & 0.27 & 0.19 & 0.14 & 0.097 & 0.069 &
0.050 \\ 
\hline
\end{tabular}
\end{center}
\caption{Production cross section for spin-0 resonance of mass
  $M_\phi$, and coupling to tops $c_\phi$=1.}
\label{tab:sigphi}
\end{table}

\section{B-tagging efficiency}
\label{sec:btag}
\begin{figure}[!htb]
\begin{center}
\includegraphics[width=0.7\textwidth]{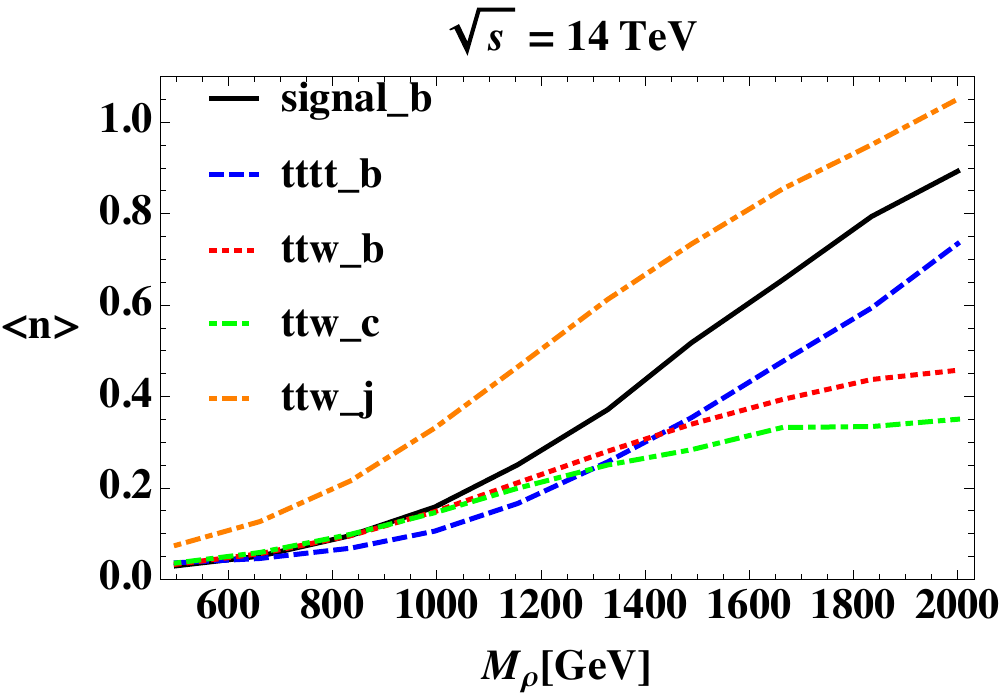}
\end{center}
\caption{The average number of b-jets,  c-jets, light-jets  with $p_T > 450$ GeV after the $H_T$ cut for the signal and the background as a function of the  mass  $M_{\rho}$. We have set the minimal value of $H_T$ to $M_{\rho}$ for each mass hypothesis. The black solid line and blue dashed line correspond  to the number of b-jets for the singal and the  SM four top background separately. The other three lines mean the number of b-jets (in read  dotted), c-jets(in green dotted-dashed ), light-jets(in orange dotted-dashed) for $t\bar{t} W + jets$.}
\label{fig:nbl}
\end{figure}

In this appendix, we want to make some comments on the constant b-tagging(mistagging) efficiency used in our analysis.   As is well known, the b-tagging  (c-mistagging) efficiency will decrease when the $p_T$ becomes too large ($p_T \gtrsim 450$ GeV). Although the mistagging rate for the  light jets will increase by a factor of 2, it is not relevant in our case, because the backgrounds originating from the light jets are two small. Our signature is mainly coming from the $3b, 4b$ configuration for the SM four top background and $2b1c$ for the $t\bar{t}W + jets$\footnote{We have checked that the fraction of events for $t\bar{t}W + jets$ coming from the $c$-mistagging rate is $\sim 70\%$.}. So both the signal and the background will be reduced for the large transverse momentum. To emphasize how large its impact, we plot  in Fig.~\ref{fig:nbl} the average number of b-jets, c-jets per-event\footnote{We only include the events with  $n_{b(c)}
\geq 1$ in our plots.} with $p_T > 450 \text{GeV}, |\eta| < 2.5$ after the $H_T$ cut for the signal and the main background as a function of the  of the resonance. From the figure, we can infer that for the signal, the effect of varying b-tagging efficiency is quite mild and it reduces the number of event by $\sim25\%$ for the signal with $M_{\rho} = 2$  TeV if we assume that the b-tagging efficiency go down from $70\%$ to $50\%$  when $p_T \gtrsim 450$ GeV.\footnote{What we really need to compare is the old efficiency $\epsilon_b = 70\%$ to the  average efficiency $ (1-\left<n_b\right>) \times \epsilon_b+ \left<n_b\right> \times \epsilon^\prime_b$ if $\left<n_b\right> < 1$, where $\epsilon^\prime_b = 50\%$.}  When the reduction of the backgrounds are also considered, the effects on the significance $S/\sqrt{B}$ are further going down to $\sim 20\%$.
So we conclude that  the constant b-tagging efficiency is a good
approximation 
in our analysis.

\section{The finite width effect}
\label{sec:finitewidth}
  As studied in Ref.~\cite{Pappadopulo:2014qza}, two kinds of  important effects due to the finite decay width are present in the searches of resonances. One is the distortion of the signal shape, as a consequence of the sharp falling of the PDF at large x, the other is the interference with SM 4 top background. 
Since $\rho$ is strongly interacting with right-handed top, it is usually much broader than  the other resonances. Neglecting the top mass, the decay-width-mass ratio is roughly $\Gamma_{\rho}/M_{\rho} \sim (1/8 \pi) g_{\rho}^2 \sim 0.04 g_{\rho  }^2$, which means that for $g_{\rho} \gtrsim 5$, the ratio is already larger than 1. In this case, it is questionable whether we can treat it as a particle or not. Possibly contact interactions should be studied. In this section, we will study the effect of decay width on the optimal cuts we imposed by adopting $g_{\rho} = 1, 2, 3, 4, 5$ for $M_{\rho} = 1 $ TeV and 2 TeV. Our result is evidently not conclusive, the dedicated analysis should be performed by the experimental collaborations.
Let's start from the effect due to the PDF.
The number of signal after all the selection cuts can be parametrized as:
\begin{equation}
\begin{split}
n_s(M_{\rho}, g_{\rho}) &= \sigma_0 (M_{\rho}, g_{\rho}, \Gamma(M_{\rho}, g_{\rho})) \times \epsilon(M_{\rho}, \Gamma (M_{\rho}, g_{\rho})) \times L
\end{split}
\end{equation}
where $\sigma_0$ is the cross section for the process $pp\rightarrow \rho t \bar{t} \rightarrow t \bar{t} t \bar{t}$ \footnote{We include all the diagrams in the presence of $\rho$ and neglect SM contributions.} before any cuts and $L$ is the integrated luminosity.  In general, the efficiency $\epsilon$ also depends on the finite decay widths. Things will be simplified when the resonance is narrow and  using NWA,  the coupling $g_{\rho}$ can be totally factorized as (for detail, see Ref.~\cite{Pappadopulo:2014qza} )
\begin{equation}
\begin{split}
n_s(M_{\rho}, g_{\rho}) &= g_{\rho}^2\times \sigma_0 (M_{\rho} ) \times \epsilon(M_{\rho}) \times L
\end{split}
\end{equation}
where we  neglect the finite decay width effects on the kinematics of the decay products. This is the formula we used when drawing the Fig.~\ref{fig:results}. As the decay width ratio $\Gamma_{\rho}/M_{\rho}$ becomes large, which is the case for large $g_{\rho}$, this procedure becomes less precise. In the following, we will quantify the finite width effects   by showing the two ratios:
\begin{equation}
\begin{split}
 R_1 & =\sigma_0 (M_{\rho}, g_{\rho}, \Gamma(M_{\rho}, g_{\rho}))/g_{\rho}^2 \sigma_0 (M_{\rho}, 1, \Gamma(M_{\rho}, 1)), \\
 R_2 & =  \epsilon(M_{\rho},  \Gamma (M_{\rho}, g_{\rho}))/\epsilon(M_{\rho}, \Gamma (M_{\rho}, 1))
\end{split}
\end{equation}
for the cases of $M_{\rho} = 1,2 $ TeV, $g_{\rho} =  2, 3, 4, 5$. Scanning over the two parameter space is beyond the scope of the paper.

From Table~\ref{tab:varygrho}, we can see that the total cross sections get a sizable contribution from the kinematical region, where the invariant mass of the two tops from the $\rho$ decay departs from the peak region  around $M_{\rho}$.  The relative difference  from naive scaling for the inclusive cross section is increasing from $16\%$ to $74\%$ as $g_{\rho}$ varying from 2 to 5 for $M_{\rho} = 1 $ TeV.  For $M_{\rho} = 2$ TeV, the situation gets worser, because it is probing the large $x$ of the gluon PDF, which drops faster. The point has already been discussed in Ref.~\cite{Pappadopulo:2014qza}. For the ratio $R_2$, the efficiency is  reduced  for  larger value of $g_{\rho }$ as expected. For comparison, we also show the numbers of $R_1 \times R_2$, which  really matter in reality. Although the inclusive cross section and the efficiency differ a lot from naive scaling,  the product of them seems well under control for $M_{\rho} = 1(2)$ TeV, which is within $15(25)\%$ even for $g_{\rho} = 5$. Nevertheless, our naive scaling is at least a conservative estimate for the large $g_{\rho}$.

\begin{table}[ht]
\begin{center}
\begin{tabular}{|c|c|c|c|c|}

\hline
    Couplings   &  $g_{\rho} = 2 $  & $g_{\rho}$ = 3   & $g_{\rho}$ = 4 &  $g_{\rho}$ = 5\\
  \hline
\hline
$R_1 (M_{\rho} = $ 1 TeV )        & 1.16      &  1.39   & 1.61   & 1.74  \\
\hline
$R_2(M_{\rho} = $ 1 TeV )        &  0.835 & 0.743 & 0.665 & 0.658           \\
\hline
$R_1\times R_2 (M_{\rho} = $ 1 TeV )      & 0.970 & 1.03 & 1.07 & 1.14  \\
\hline
$R_1 (M_{\rho} = $ 2 TeV )        &   2.02     & 3.41     & 4.57    &  5.08   \\
\hline
$R_2(M_{\rho} = $ 2 TeV )        &   0.511 & 0.313 & 0.261 & 0.240   \\
\hline
$R_1\times R_2 (M_{\rho} = $ 2 TeV )   & 1.03 & 1.07 & 1.19 & 1.22 \\
\hline
\end{tabular}
\end{center}
\caption{Relative efficiencies after all the selection cuts  under  the different  couplings of the $\rho$ resonance.}
\label{tab:varygrho}
\end{table}

As regards with the inteference with SM four top background, we  have calculated the total  cross section  including  the interference terms and compare them with direct sum of the cross sections. It turn out that the interference effects are well under control in our case and rarely exceed 10\%. This can be due to the fact that the relevance of  the interference term is dertermined by the two competing effects: the decay width and  the ratio between the signal and the four top background. The larger the decay width and the smaller the signal to background ratio, the more important for the interference contribution. But in our case, both of them are fixed by the same parameter $g_{\rho}$ and have the same scaling  $\sim g_{\rho}^2$, which cancelled with each other and   resulted in the quite mild behaviour for the interference term.

\section{Statistical tools}
\label{sec:stats}
To obtain our final results, following \cite{Contino:2012xk} we define
a Bayesian posterior probability $p_\mathcal{L}(\sigma|N_\text{obs})$ of a total event
cross section, $\sigma$, given an observed number of events, $N_\text{obs}$, at
an integrated luminosity, $\mathcal{L}$,
as the product
of a Poissonian likelihood function $L(N_\text{obs}|\sigma\mathcal{L})$ and a prior
$\pi(\sigma)$:
\begin{equation}
p_\mathcal{L}(\sigma|N_\text{obs})\propto L(N_\text{obs}|\,\sigma\mathcal{L})\;\pi(\sigma)
\end{equation}
where
\begin{equation}
L(N|\,\sigma\mathcal{L})=\frac{\exp^{-\sigma\mathcal{L}}\left(\sigma\mathcal{L}\right)^{N_\text{obs}}}{N_\text{obs}!}
\end{equation}
is the poissonian probability of observing $N_\text{obs}$ events, with a given
process cross-section and integrated luminosity.  

In order to obtain the discovery contours of
Figs.~\ref{fig:results}(a) and (d), we take a prior that
is flat for all $r>0$, and vanishing otherwise, and normalize the probability
such that
\begin{equation}
\int_0^{\infty}d\sigma\; p_{\mathcal{L}}(\sigma|N)=1
\end{equation}
We then compute, at each point in the
($m_\rho,g_\rho$) parameter space, corresponding to a given signal and
background cross-section ($\sigma_S$ and $\sigma_B$), the smallest luminosity at
which there are more than 5 observed events
$(\sigma_S+\sigma_B)\mathcal{L}\geq 5$,
and the following inequality is satisfied:

\begin{equation}
\int_0^{\sigma_B}d\sigma'\,p_{\mathcal{L}}\!\left(\sigma'\big{|}
(\sigma_S+\sigma_B)\mathcal{L}\right)
\leq 5.7\times 10^{-7}.
\end{equation}
This corresponds to the possibility of a cross section smaller than or equal to that of
the background being consistent with a measured total number
$(\sigma_S+\sigma_B)\mathcal{L}$ events occuring less than $5\times 10^{-5}\%$
of the time (=5$\sigma$ in the large statistics limit).

To obtain the parameter measurement plot in Fig.~\ref{fig:results}(b) we
normalize the posterior probability independently at each resonance
mass, with a prior distribution that is flat over the range of
couplings $g_\rho$ in $(0,4\pi)$ as $\int dg_\rho\; p_\mathcal{L}\!\left(\sigma(M_\rho,g_\rho)\big|N_\text{obs}\right)=1$
and compute the value of the coupling at which the posterior
probability with injection of the SM contained within the region is 5\%.
Note that this procedure is sensitive to the choice of
prior, if the boundary is placed in a region where the probability is
changing rapidly.  

To obtain the 95\% upper limit on the cross section, we follow the
procedure above, except we normalize the posterior
probability with a flat prior over the range $\sigma$ in
$(0,\infty)$.  Note, however, that the appropriate lower limit will
depend on the model in question; in
the case of the $SO(5)/SO(4)$ Composite Higgs model, for example,
$g_{\rho_X}$ must be larger than the SM hypercharge coupling
$g^\prime$.  This is a consequence of the same prior-dependence noted
above. The result is much less sensitive to the choice of upper
limit, since the posterior probability for much of the range of large
$\sigma$ is negligible.

\section{Vector singlet in $SO(5)/SO(4)$ Composite Higgs model}
\label{sec:so5modso4}
We briefly review the properties of the $\rho_X$ composite vector singlet
in the $SO(5)/SO(4)$ CH model below.For a detailed exposition and analysis,
see~\cite{Greco:2014aza,Contino:2011np}. In the limit $M_*\gg M_{\rho_X}$,
where $M_*$ is the mass scale of all the other bounds states of the
strong sector, we can integrate out all other heavy resonances,
giving, at leading order in the derivative expansion, the following
effective lagrangian:  
\begin{equation}
\label{Lag}
\begin{split}
\mathcal{L} =  
&\quad -\frac{1}{4} W_{\mu\nu}^a W^{a\, \mu\nu} -\frac{1}{4} B_{\mu\nu} B^{\mu\nu}
+ \bar{\psi} \gamma^\mu \!\left(i \partial_\mu + g_{el} \frac{\sigma^a}{2} W_\mu^a  P_L + Y g_{el}^\prime  B_\mu \right)\! \psi + \frac{f^2}{4} (d_\mu^{\hat{a}})^2 \\
& - \frac{1}{4} \rho_{X\mu\nu} \rho^{\mu\nu}_X+ \frac{M_{\rho_X}^2}{2 g_{\rho_X}^2} (g_{\rho_X} \rho_{X\mu} - g_{el}^\prime B_\mu)^2\, + c \,\bar{t}_R\gamma^\mu(g_{\rho_X} \rho_\mu - g_{el}^\prime B_\mu) t_R,  \\ 
\end{split}
\end{equation}
where $g_{el}$ are the proto-electroweak gauge couplings, $c$ is an
$\mathcal{O}(1)$ parameter and $\psi$ stands for all the SM
fermions.\footnote{ Note that
there is a linear mixing term between $\rho_X^\mu$ and $B^\mu$ before
electroweak symmetry breaking (EWSB), since only the difference
$g_{\rho_X}\rho_X^\mu-g^\prime_{el}B^\mu$ is invariant under the $U(1)_X$ symmetry.}. Here we assume that the RH top quark is a chiral singlet bound
state of the strong sector, which allows it to couple directly to  
$\rho_X$ as shown above. $d_\mu^{\hat{a}}$ is defined via the CCWZ
construction as a function of the $SO(5)/SO(4)$ Nambu-Goldstone matrix $U$:
\begin{align}
\label{eq:dEdef}
- i U^\dagger D_\mu U = d_\mu + E_\mu\, ,
\end{align}
where $U =  \text{exp}(i\sqrt{2}\pi^{\hat{a}} T^{\hat{a}}/f)$.  Under
a general $SO(5)$ rotation $g\,\in \,SO(5)$,  this is subject to
the unbroken $SO(4)$ transformation as follows:
\begin{equation}
U \rightarrow g\, U\, h(x)^\dagger, \qquad d_\mu \rightarrow h(x)
\,d_\mu \,h(x)^\dagger, \qquad E_\mu \rightarrow h(x)\, E_\mu\,
h(x)^\dagger - i h(x)\, \partial_\mu h(x)^\dagger
\end{equation}
where $h(x) \in SO(4)$ is a complicated function of $(\pi(x), g)$.
Going to unitary gauge after EWSB:

\begin{align}
\label{dEuni}
d_\mu^{\hat{a}} &= -\frac{\sin(\theta + h/f)}{\sqrt{2}}\delta^{\hat{a} \hat{i}}\left(g_{el} W_\mu^{\hat{i}} - g^\prime_{el} \delta^{\hat{i}3}B_\mu\right)+ \sqrt{2} \frac{\partial_\mu h}{f} \delta^{\hat{a} 4} , \nonumber \\
E_\mu^{aL} & = -\left(\frac{1 + \cos(\theta + h/f)}{2}\right) g_{el} W_\mu^a - \delta^{a3}\left(\frac{1 - \cos (\theta + h/f)}{2}\right) g^\prime_{el} B_\mu , \nonumber\\
E_\mu^{aR} & = - \left(\frac{1 - \cos(\theta + h/f)}{2}\right) g_{el} W_\mu^a - \delta^{a3}\left(\frac{1 + \cos(\theta + h/f)}{2}\right) g^\prime_{el} B_\mu \,.
\end{align}
where $ \hat{i}=1\cdots 3$ and $\theta = \left< h\right> /f$ is the
vacuum misalignment angle, which can be treated as an order parameter for the EWSB. The $W$ mass is easily obtained by using above expressions, which gives  $m_W^2 = \frac14 g_{el}^2 f^2\sin^2\theta$. One can identify the $SU(2)_L$ gauge coupling and the usual  EWSB scale $g = g_{el}$ and $v = f \sin\theta$. For neutral spin-1 sector, the mass matrix after EWSB  is straightforward to obtain :
\begin{equation}
M_{\rho_X^{0}}^2 =\left(
\begin{array}{ccc}
\frac{g_{el}^2 f^2 \sin^2\theta}{4}
& 0
& - \frac{g_{el}  g_{el}^\prime f^2 \sin^2\theta}{4}  \\
 0
& m_{\rho_X}^2 
& - \frac{   g_{el}^\prime m_{\rho_X}^2
}{ g_{\rho_X}} \\
-\frac{g_{el}  g_{el}^\prime f^2 \sin^2\theta}{4} 
& - \frac{   g_{el}^\prime m_{\rho_X}^2
}{ g_{\rho_X}}
& \frac{g_{el}^{\prime2} f^2 \sin^2\theta}{4}  +  \frac{   g_{el}^{\prime2} m_{\rho_X}^2
}{ g_{\rho_X}^2}
 \\
\end{array}
\right)
\end{equation}
Using the expression for the $m_W^2$, we can rewrite the mass matrix as follows:

\begin{equation}
M_{\rho_X^{0}}^2 =m_{\rho_X}^2 \left(
\begin{array}{ccc}
m_W^2/m_{\rho_X}^2 & 0  & - (m_W^2/m_{\rho_X}^2)  g_{el}^\prime/g \\
 0 &1  & -  g_{el}^\prime / g_{\rho_X} \\
- (m_W^2/m_{\rho_X}^2)  g_{el}^\prime/g  & -  g_{el}^\prime / g_{\rho_X} \
& g_{el}^{\prime2}/g_{\rho_X}^2 +  (m_W^2/m_{\rho_X}^2) g_{el}^{\prime2}/g^2   
 \\
\end{array}
\right)
\end{equation}
from which we immediately notice that the true small expansion parameter
in the mass matrix  is $m_W^2/m_{\rho_X}^2$. The physical masses of
the $\rho_X$ and $Z$ boson are obtained by diagonalizing the mass
matrix at linear order in $m_W^2/m_{\rho_X}^2$ :
\begin{equation}
\begin{split}
M_{\rho_X} &= \frac{m_{\rho_X}}{\sqrt{1 - g^{\prime 2}/g_{\rho_X}^2}} \left( 1 +\frac12 \frac{g^{\prime 4}}{g^2 g_{\rho_X}^2}\frac{m_W^2}{m_{\rho_X}^2}\right)\\
m_Z & = \frac{1}{2}\sqrt{g^{\prime2} + g^2}\,v
\end{split}
\end{equation}
for $g^{\prime -2}=g_{el}^{\prime -2}+g_{\rho_X}^{-2}$.
Are rotating to the mass eigenstates, we can obtain the interactions between the $\rho_X$ and  SM particles, which are parametrized as
follows (in the conventions of~\cite{Greco:2014aza}):
\begin{equation}
\label{eq:cubic}
\begin{split}
\mathcal{L}_{\rho_X}   =  \, 
& i g_{\rho_X W W}\, \big[  (\partial_\mu W_\nu^+ - \partial_\nu W_\mu^+ ) W^{\mu -}\rho^{\nu}_X  
                                  + \frac 12 (\partial_\mu \rho_{X\nu} - \partial_\nu \rho_{X\mu} ) W^{\mu +}W^{\nu -}  + h.c.\big]  \\[0.15cm] 
&+  g_{\rho_X Z h}  \, h  \rho_{X\mu} Z^{\mu}   
+  \rho_{X\mu}\, \bar{\psi}_u \gamma^\mu \!\left[ \frac 12 (g_{\rho_X ffL} - g_{\rho_X ffY} ) P_L+ g_{\rho_X ffY } Q[ \psi_u] \right]\! \psi_u  \\[0.15cm]
& + \rho_{X\mu} \,\bar{\psi}_d \gamma^\mu  \!\left[ -\frac 12 (g_{\rho_X ffL} - g_{\rho_X ffY} )P_L + g_{\rho_X ffY } Q[ \psi_d] \right] \!\psi_d \, ,
\end{split}
\end{equation}
where $\psi_u$ ($\psi_d$) stands for any of the SM up-type quarks and neutrinos
(down-type quarks and charged leptons). 
The couplings are given by:

\begin{equation}
\label{eq:couplings}
\begin{split}
g_{\rho_X WW} &=  \frac{g^{\prime 2}}{g_{\rho_X}} \frac{m_W^2}{M_{\rho_X}^2}\frac{1}{\sqrt{1 - g^{\prime 2}/g_{\rho_X}^2} },\qquad g_{\rho_XZh} =  \frac{g^{\prime 2}}{g_{\rho_X}} \frac{m_Z}{\sqrt{1 - g^{\prime 2}/g_{\rho_X}^2}}\\
g_{\rho_XffL}&=  \frac{g^{\prime 2}}{g_{\rho_X}} \frac{m_W^2}{M_{\rho_X}^2}\frac{1}{\sqrt{1- g^{\prime 2}/g_{\rho_X}^2}}\\
g_{\rho_XffY}&= - \frac{g^{\prime 2}}{g_{\rho_X}} \frac{1}{\sqrt{1 - g^{\prime 2}/g_{\rho_X}^2}} -
\frac{g^{\prime 2}}{g_{\rho_X}} \frac{g^{\prime 2}}{g^2} \frac{m_W^2}{M_{\rho_X}^2} \frac{1}{\sqrt{1 - g^{\prime 2}/g_{\rho_X}^2}} \\
g_{\rho_XttL}&= \frac 12 g_{\rho_XffL} + \frac16 g_{\rho_XffY}\\
g_{\rho_XttR}&=  c  \frac{g_{\rho_X} }{\sqrt{1 - g^{\prime 2}/g_{\rho_X}^2}} + \frac23 g_{\rho_XffY}\\
\end{split}
\end{equation}
where we have substituted the identity:
\begin{equation}
m_{\rho_X} = M_{\rho_X} \sqrt{1 - g^{\prime 2}/g_{\rho_X}^2}.
\end{equation}
We can see that the coupling of $\rho_X t_L \bar{t}_L$ is suppressed by  a
 factor of $g^{\prime 2}/(6g_{\rho_X}^2$) compared to $\rho_X t_R
 \bar{t}_R$. In the high energy limit, the cross section for $t\bar{t}$ fusion to $\rho_X$ will be proportional to $g_{\rho_XttL}^2 + g_{\rho_XttR}^2$, so in most of the case, the coupling to left-handed top can be neglected.
 Note that for the
couplings and the masses of the $\rho_X$, there is a univeral factor
of  $1/\sqrt{1 - g^{\prime 2}/g_{\rho_X}^2}$ from the difference of $g_{el}^\prime$ and $g^\prime$. Unless we consider extremely small $g_{\rho_X} \sim g^\prime$, in which case that this factor is $\mathcal{O}(1)$, our expansion in $m_W^2/m_{\rho_X}^2$ is safe. Actually, both small $g_{\rho_X}$ and small $m_{\rho_X}$ is also excluded by the $Y$ parameter constraint:
\begin{equation}
 g_{\rho_X} m_{\rho_X} \geq 836 \text{GeV}.
\end{equation}

Concerning the decay of $\rho_X$, the relevant modes are $WW, Zh, f\bar{f}$, where $f$ denotes the SM chiral fermions. For the fully elementary SM fermions, the couplings to $\rho_X$ are universal and  the decays into  them are purely determined by the two form factors $g_{\rho_XffL}, g_{\rho_XffY}$ defined in Eq.~\ref{eq:cubic}. We can see from Eq.~\ref{eq:couplings} that $g_{\rho_X ffL}$ is suppressed by a factor of $m_W^2/m_{\rho_X}^2$ and can be safely neglected. We present here the analytical formulae for the decay widths in the large coupling limit and neglect all the masses of the SM particles :

\begin{equation}
\begin{split}
\Gamma(\rho^0_X \rightarrow W^+ W^-)/M_{\rho_X} &=\frac{g^{\prime 4}}{192\pi  g_{\rho_X}^2}  \\ 
\Gamma(\rho^0_X \rightarrow Z h) /M_{\rho_X}&= \frac{g^{\prime 4}}{192\pi g^2_{\rho_X}}  \\
\Gamma(\rho^0_X \rightarrow \psi_f \bar{\psi}_f )/M_{\rho_X} &=\frac{N_f^c Y_f^2 g^{\prime 4}}{24\pi g^2_{\rho_X}}   \\
\Gamma(\rho^0_X \rightarrow t_R \bar{t}_R )/M_{\rho_X} &=\frac{ c^2 g^2_{\rho_X}}{8\pi }   \\
\end{split}
\end{equation}
where $Y_f$ is the hyper-charge for the elementary chiral fermions  in
SM and $N_f^c$ denotes the color factor of the fermions. Note that the decay width to gauge bosons are suppressed by a
kinematical factor of 8 compared with that of the fermions, which
makes the channels are less important. We can also see that for the
fully composite $t_R$, the ratio of  branching fraction of top pair to
that of elementary fermions scales as $g_{\rho_X}^4/g^{\prime 4}$.

\newpage

\end{document}